\documentclass[oneside,reqno]{amsart}

\usepackage{cite}

\usepackage{geometry}

\usepackage{newtxtext}
\usepackage{babel}

\usepackage[active]{srcltx}
\usepackage{xcolor}

\usepackage{amstext}
\usepackage{amsthm}
\usepackage{amssymb}

\usepackage{hyperref}
\usepackage{graphicx}

\usepackage{pgfplots}
\pgfplotsset{compat=1.16} 
\usepackage{caption}

\makeatletter

\numberwithin{equation}{section}
\numberwithin{figure}{section}
\theoremstyle{remark}
\newtheorem{remark}{Remark}[section]

\theoremstyle{plain}
\newtheorem{theorem}{Theorem}[section]
\newtheorem{lemma}{Lemma}[section]

\newtheorem{proposition}{Proposition}[section]

\allowdisplaybreaks

\DeclareMathOperator{\Tr}{Tr}

\DeclareMathOperator{\ImPart}{Im}

\DeclareMathOperator{\sign}{sign}

\newcommand{\bR}{\mathbb{R}}
\newcommand{\bN}{\mathbb{N}}

\newcommand{\cH}{\mathcal{H}}
\newcommand{\cL}{\mathcal{L}}
\newcommand{\cN }{\mathcal{N}}
\newcommand{\cU}{\mathcal{U}}
\newcommand{\cR}{\mathcal{R}}
\newcommand{\cF}{\mathcal{F}}

\newcommand{\ii}{\mathrm{i}}
\newcommand{\ri}{\mathrm{i}}
\newcommand{\dx}{\mathrm{d}x}
\newcommand{\dy}{\mathrm{d}y}

\newcommand{\dZ}{\mathrm{d}Z}
\newcommand{\ds}{\mathrm{d}s}
\newcommand{\dt}{\mathrm{d}t}

\newcommand{\dr}[1]{\mathrm{d}{#1}\, }

\usepackage{dsfont}

\makeatother

\begin{document}
	\title[uniform convergence to BEC with an external potential]{Uniform in Time Convergence to Bose--Einstein Condensation for a Weakly Interacting Bose Gas with an External Potential}
	
	\author[C. Dietze]{Charlotte Dietze}
	\address[C. Dietze]{Department of Mathematics, 
		LMU Munich, Theresienstraße 39,
		80333 Munich,
		Germany}
	\email{dietze@math.lmu.de}
	
	\author[J. Lee]{Jinyeop Lee}
	\address[J. Lee]{Department of Mathematics, 
		LMU Munich, Theresienstraße 39,
		80333 Munich,
		Germany}
	\email{lee@math.lmu.de}

	\begin{abstract}
		We consider a gas of weakly interacting bosons in three dimensions subject to an external potential in the mean field regime. Assuming that the initial state of our system is a product state, we show that in the trace topology of one-body density matrices, the dynamics of the system can be described by the solution to the corresponding Hartree type equation. Using a dispersive estimate for the Hartree type equation, we obtain an error term that is uniform in time. Moreover, the dependence of the error term on the particle number is optimal.
	\end{abstract}
	
	\maketitle

\section{Introduction}\label{sec:intro}

Bose-Einstein condensation was proposed by Bose and Einstein \cite{bose,ei1925} in 1924.
Since then, the topic has gained great interest in both physics and mathematics \cite{gross,pitaevskii,leggett,pomeaurica,benedikter2016effective,lieb2005mathematics}, in particular after the first experimental observation by  Wieman, Cornell and Ketterle \cite{wiemancornell,ketterle} in 1995. 

We consider the dynamics of a Bose gas of $N$ particles in three
dimensions interacting through a symmetric two-body potential in the presence of 
an external potential.
We assume our initial state $\psi_{N,0}\in L^2(\bR^{3N})$ to be fully factorised, i.e.,
\begin{equation}\label{eq:factorized-initial}
	\psi_{N,0} (x_1, \dots, x_N) = \prod_{j=1}^N \varphi_0(x_j)
\end{equation}
for some $\varphi_0\in L^2(\bR^3)$ with $\|\varphi_0\|_2=1$.
The time evolution of our state is governed by the Hamiltonian
\begin{equation}\label{eq:N_body_Hamiltonian}
	H_{N}=\sum_{j=1}^{N}\big(-\Delta_{x_{j}}+V(x_{j})\big)+\frac{\lambda}{N}\sum_{i<j}^{N}w(x_{i}-x_{j}).
\end{equation}
The time-evolution $\psi_{N,t}$ is the solution to 
\begin{equation}
	\begin{cases}
		\ii \partial_t \psi_{N,t} &= H_{N} \psi_{N,t}\\
		\left.\psi_{N,t}\right|_{t=0} &= \psi_{N,0}.
	\end{cases}
\end{equation}
Alternatively, one can write
\[
\psi_{N,t} = e^{-\ii H_N t} \psi_{N,0}
\]
for every $t\in \bR$.
Here, $V:\bR^3\to\bR$ denotes the external potential and
the interaction potential $w\in L^1(\bR^3) \cap L^2(\bR^2)$ is independent of $N$.
We consider weakly interacting bosons, so the coupling constant $\lambda\in\bR$ is small in the sense that $|\lambda|\leq \lambda_0$ for some $\lambda_0$ depending on $\varphi_0$, $\|w\|_1$ and $V$.
We will give the precise assumptions on the external potential $V$, on $\varphi_0$, and on $\lambda_0$ later.

We prepare a product state that is close to the `ground state of an interacting Bose gas in a trap', which is called the Bose-Einstein condensation state. Then we want to investigate the time evolution of the state after removing the trap and replacing it by a small external potential $V$.

Since the initial state $\psi_{N,0}$ is a product state, see \eqref{eq:factorized-initial}, we expect that, in an appropriate sense, 
the time-evolved state $\psi_{N,t}$ is also approximately given by a product state $\prod_{j=1}^N \varphi_t(x_j)$ for some $\varphi_t\in L^2(\bR^3)$, where $\varphi_t$ is the solution to the corresponding 
Hartree-type equation \eqref{eq:Hartree}
\begin{align}
	&\begin{cases}
		\ii \partial_t \varphi_t &= (-\Delta +V) \varphi_t + \lambda (w * |\phi_t|^2) \varphi_t\\
		\left.\varphi_t\right|_{t=0} &= \varphi_0.
	\end{cases}\tag{Hartree}\label{eq:Hartree}
\end{align}

We cannot expect $\prod_{j=1}^N \varphi_t (x_j)$ and $\psi_{N,t}$ to be close in the $L^2$-norm sense, see \cite{LewinNamSchlein15}. However, we can prove that they are close in the trace topology of one-body density matrices as $N\to \infty$. More precisely, we define the marginal one-particle density $\gamma_{N,t}^{(1)}$ by its operator kernel
\begin{equation}\label{eq:Kernel_of_Marginal_Density}
	\gamma_{N,t}^{(1)}(x;y) = \int_{\bR^{3(N-1)}} \dZ \; \overline{\psi_{N,t}(y,Z)}\,\psi_{N,t}(x,Z).
\end{equation}

Our main result is the following.
\begin{theorem}
	\label{thm:main}
	Let $N\in \bN$. 
	Suppose that $w$ is even, real-valued, and $w\in L^{1}(\bR^{3})\cap L^{2}(\bR^{3})$.
	Let $V\in W^{2,\infty}(\bR^3)$ be real-valued and such that there exists a constant $C^V>0$ such that
	\[
	\|e^{\ii t (-\Delta + V)} f\|_\infty \leq C^V \|f\|_1
	\]
	for all $f\in L^1(\bR^3)\cap L^2(\bR^3)$.
	Let $\varphi_0\in H^{2}(\bR^3)$ with $\|\varphi_0\|_2 = 1$. 
	We let our system have fully factorized initial data, i.e.,
	\[
	\psi_{N,0}(\mathbf{x})=\prod_{j=1}^{N}\varphi_0(x_{j})
	\]
	for $\mathbf{x}=(x_{1},x_{2},\dots,x_{N})\in\bR^{3N}$.
	Let
	\[
	\psi_{N,t} = e^{-\ii H_N t}\psi_{N,0}
	\]
	for $t>0$ where $H_N$ is defined in \eqref{eq:N_body_Hamiltonian}. 
	Here, we assume that $|\lambda|\leq \lambda_0$ with $0<\lambda_0\le1$ depending only on $\|V\|_{W^{2,\infty}}$, $C^V$, $\|w\|_1$, $\|\varphi_0\|_{H^2}$, and $\|e^{\ii (-\Delta + V) } \varphi_0 \|_1 $.
	Then there exists a constant $C>0$ depending only on 
    $\|V\|_{W^{2,\infty}}$, $C^V$, $\|w\|_1$, $\|w\|_2$, $\|\varphi_0\|_{H^2}$, and $\|e^{\ii (-\Delta + V) } \varphi_0 \|_1 $
	such that
	\begin{equation}
		\label{eq:main}	\Tr\Big|\gamma_{N,t}^{(1)}-|\varphi_t\rangle\langle \varphi_t|\Big|\leq C N^{-1}
    \end{equation}
	for all $t>0$.
	In particular, the constant $C$ does \emph{not} depend on $t$ or $N$.
\end{theorem}

\begin{remark}\phantom{ }
	\begin{enumerate}
		\item
		A special case of Theorem \ref{thm:main}, namely without external potential, i.e.  $V=0$, was proved in \cite[Theorem 1.3]{Lee2019time}. Theorem \ref{thm:main} is new because we can also treat external potentials.
		\item For any $\lambda\geq0$ with $w\in C^1(\bR^3)$ vanishing at infinity, non-negative, spherically symmetric, and decreasing, one can prove the same result without external potential by using \cite[Corollary 3.4]{GrillakisMachedon13}. In particular, we can cover large $\lambda>0$.
		\item
		In many occasions, relevant physical quantities can be obtained from the one-particle density matrix, so Theorem \ref{thm:main} tells us that in these occasions we can obtain a good approximation by replacing the one particle density $\gamma_{N,t}^{(1)}$ by $| \varphi_t\rangle \langle \varphi_t|$.
		
		\item
		A similar theorem can be obtained for general $\lambda$ (without smallness assumption) with time dependent constant $C$.

	\end{enumerate}
\end{remark}

Our proof is based on \cite{RodnianskiSchlein2009,ChenLeeSchlein2011,ChenLeeLee2018}.
We derive the large $N$ limit with interaction potential $w\in L^{1}(\bR^{3})\cap L^{2}(\bR^{3})$
under the existence of external potential $V$. 

We track the dependence of $\|\varphi_{t}\|_{\infty}$, which helps us get a better estimate after using the Grönwall inequality. Applying the time decay estimate obtained in \cite{Dietze2021},
we obtain a time independent rate of convergence.
We will explain more details of the proof strategy later in Section \ref{sec:proof-strategy}.

\subsection{History}

\subsubsection*{Many-Body Convergence}

Now we are going to review the results on the derivation of the effective one-body Schrödinger equation from the many-body Schrödinger equation. To be more precise, a typical result in this direction is that if one has an initial many-body wave function $\psi_{N,0}$ with
\begin{equation}
	\lim_{N\to\infty} \Tr \left| \frac{1}{N} \gamma_{N,0} - |\varphi_0\rangle\langle \varphi_0| \right| = 0
\end{equation}
for some given $\varphi_0\in L^2(\bR^3)$, then at later times $t>0$,
\begin{equation}\label{eq:convergence}
	\lim_{N\to\infty} \Tr \left| \frac{1}{N} \gamma_{N,t} - |\varphi_t\rangle\langle \varphi_t| \right|=0,
\end{equation}
where $\varphi_t$ is a solution of an effective partial differential equation.

For the mean-field regime, with differentiable interaction potential $w$, Hepp \cite{Hepp1974} showed \eqref{eq:convergence}. Moreover, Spohn \cite{Spohn1980} obtained \eqref{eq:convergence} for a bounded interaction potential $w$. Ginibre and Velo provided a series of works in this direction \cite{GinibreVelo1979_1,GinibreVelo1979_2,GinibreVelo1998}.

Later, for singular potentials, the derivation of \eqref{eq:convergence} was obtained by Bardos, Golse, and Mauer \cite{BardosGolseMauser2000} for $L^2(\bR^3)+L^\infty(\bR^3)$-potentials in one dimension, and by Erd\H{o}s and Yau \cite{ErdosYau2001} for the Coulomb potential in three dimensions. These works are based on the BBGKY\footnote{Bogoliubov–Born-Green–Kirkwood–Yvon} hierarchy method. Therefore, the rate of convergence in these works was not given explicitly.
Moreover, the convergence is only established for any time of order one.

For factorized initial states,
Rodnianski and Schlein \cite{RodnianskiSchlein2009} developed a scheme to obtain an explicit rate of convergence. They showed
\begin{equation}
	\Tr \left| \frac{1}{N} \gamma_{N,{t}} - |\varphi_{t}\rangle\langle \varphi_{t}| \right| \leq \frac{Ce^{Ct}}{\sqrt{N}}
\end{equation}
for interaction potentials $w$ including the Coulomb interaction potential.
Knowles and Pickl \cite{KnowlesPickl2010} provided a scheme to cover more singular potentials than Coulomb with the same rate based on the techniques in \cite{Pickl2011}.

For bounded and integrable interaction potentials, the optimal rate in $N$ was provided in \cite{ErdosSchlein2009}. This work was extended in \cite{ChenLee2011}. In \cite{ChenLeeSchlein2011}, Chen, Lee, and Schlein provided the rate $Ce^{Ct}/N$, which is optimal in $N$, for the Coulomb interaction potential. This work was generalized in \cite{ChenLeeLee2018} to more singular interaction potentials using Strichartz estimates for the Hartee equation. The time dependence was investigated in \cite{Lee2019time}.

Elgart, Schlein \cite{ElgartSchlein2007} and Michelangeli, Schlein \cite{MichelangeliSchlein2012} proved convergence for the mean-field Boson star equation, i.e.,
\[
H_{N}=\sum_{j=1}^{N}\big(1-\Delta_{x_{j}}\big)^{1/2}+\frac{\lambda}{N}\sum_{i<j}^{N}\frac{1}{x_{i}-x_{j}}.
\]
Later Lee provided the rate of convergence \cite{Lee2013}.

For the intermediate regime and GP regime,
Erd\H{o}s, Yau, and Schlein \cite{ErdosSchleinYau2007,ErdosSchleinYau2009} proved \eqref{eq:convergence}.
Moreover, for the GP regime, Benedikter, de Oliveira and Schlein \cite{NielsOliveiraSchlein2015}, Pickl \cite{Pickl2015}, and Brennecke and Schlein \cite{BrenneckeSchlein2019GP} provided explicit convergence rates.

It is important to have a time independent bound to show the stability of Bose-Einstein condensation.
There have been a few works in the direction of improving time dependence.
In \cite{GrillakisMachedonMargetis2011}, the authors provided a norm approximation with a bound of the form $C(1+t)^{1/2}/N$ for a cut-offed Coulomb interaction potential.
In \cite{kuz2015rate}, with a cut-offed Coulomb potential, the trace norm was bounded by $C(1+t)/N$. 
In \cite{Lee2019time}, for interaction potentials of the form $\lambda e^{-\mu|x| }|x|^{-\gamma}$ without cut-off, trace norm bounds have been proved. {For the detailed time-dependent behaviour of the error, see, \cite[Table 1]{Lee2019time}.}
Here, in this paper, we provide a time independent bound of the trace norm by using the time decay estimate for one-body nonlinear Schr\"odinger equations.

The second-order correction was obtained by a norm approximation by a series of works by Grillakis, Machedon and  Margetis \cite{GrillakisMachedon13,GrillakisMachedon17,GrillakisMachedonMargetis2010,GrillakisMachedonMargetis2011}. 
Pickl \cite{pickl2008gphtr} also considered this regime.
Recently, Napiórkowski provided a nice review paper about this topic \cite{napiorkowski2021dynamics}.
There are many directions to investigate further. For example, there are at least two directions. Regarding mixture condensation, one can find results in \cite{michelangeli2019ground,michelangeli2017mean,de2019mean,Lee2021mixture,dimonte2020some}. Results on three-body interactions can be found in \cite{nam2021condensation,lee2020rate,chen2011quintic,chen2012second} and in the references therein.
There are also several results \cite{boccato2017quantum,brennecke2019fluctuations,LewinNamSchlein15,nana2015, namnapiorkowski,GrillakisMachedonMargetis2010,GrillakisMachedonMargetis2011,GrillakisMachedon13,GrillakisMachedon17}
on a norm approximation, which is much more precise than the approximation of the one particle marginal density in trace norm. As we have already mentioned earlier, it is \emph{not} 
possible to show that $\psi_{N,t}$ is close to a product state in the norm sense, and instead, one has to take excitations into account. This can be done by using Bogoliubov theory, which requires a more detailed analysis of the Hamiltonian in Fock space than the method used in this paper.

{In \cite{ChenLeeSchlein2011}, they mentioned in a remark that their approach could cover external potentials. This was tracked in \cite{leemichelangeli2022}.}
Other results including external potentials are  \cite{digi2021, lenaro2014, brscsc2022, nanaritr2020, brscsc20222, natr2021}.

Since there is a correlation among particles, it might not be natural to expect a time independent bound for the norm approximation.
In this paper, we provide the convergence in terms of a trace norm approximation of one-particle density matrices instead of a norm approximation of wave functions.

\subsubsection*{Time Decay Estimate for One-Body Nonlinear Schrödinger Equations }

Let us first review several results on one-body nonlinear Schrödinger equations without external potentials. Ginibre and Velo \cite[Theorem 6.1(1)]{ginibrevelononlocal} considered Hartree type nonlinearities with repulsive interaction potentials $w$ and they proved the decay estimate
\begin{equation}
	\| \varphi_t\|_q\le C(1+|t|)^{-\frac{d}{2}\big(\frac{1}{q'}-\frac{1}{q}\big)}\ \textrm{for all } t\neq0
\end{equation}  
for all $q$ with $\big[\frac{1}{2}-\frac{1}{d}\big]_+\le\frac{1}{q}\le\frac{1}{2}$, where $[\;\cdot\;]_+$ denotes the positive part.
Note that $\big(\frac{1}{2}-\frac{1}{d}\big)^{-1}=\big(\frac{d-2}{2d}\big)^{-1}=\frac{2d}{d-2}$, so their result corresponds to the energy critical and energy subcritical case. In particular, the case $q=\infty$ is not covered in dimension $d=3$. Hayashi and Naumkin \cite{HayashiNaumkin98} looked at critical nonlinearities, namely the local nonlinearity 
\begin{equation}
	\lambda | \varphi_t|^{\frac{2}{d}} \varphi_t \quad \textrm{for}\quad \lambda\in\bR\ \textrm{and  } d\in\{1,2,3\}
\end{equation} 
and the nonlocal nonlinearity 
\begin{equation}
	\lambda (|\cdot|^{-1}*| \varphi_t|^2) \varphi_t \quad \textrm{for} \quad \lambda\in\bR\ \textrm{and  } d\ge2
\end{equation}
and they can also add non-critical nonlinearities. They showed a dispersive estimate of the form
\begin{equation}\label{eq:infdec}
	\| \varphi_t\|_\infty\le C(1+|t|)^{-\frac{d}{2}}\ \textrm{for all } t\neq0\,.
\end{equation} 
Later Kato and Pusateri \cite{katopusateri} gave an alternative  proof of the result in \cite{HayashiNaumkin98}, which was based on a careful analysis of the equation in Fourier space.

Grillakis and Machedon \cite[Corollary 3.4]{GrillakisMachedon13} considered Hartree type nonlinearities with nonnegative, radial, decreasing interaction potential $w\in L^1(\bR^d)\cap C^1_0(\bR^d)$ for sufficiently regular but possibly large initial data. They showed a decay estimate of the form \eqref{eq:infdec} using a Grönwall type argument after proving suitable a-priority estimates. Their result was applied in \cite{namnapiorkowski}, where they showed a norm approximation for the dynamics of many-body quantum systems in the context of Bose-Einstein condensation. Other results without external potentials we would like to mention are \cite{ginibrevelotimedecaykleingordonnls, hayashitsutsumi, cazenaveweissler, ozawa, ginibreozawa, ginibreveloscatteringhartree, deiftzhou, duyckaertsholmerroudenko, choozawa}.

Next, let us turn to results on nonlinear
Schrödinger equations with external potentials.
In dimension $d=1$, Cuccagna, Georgiev and Visciglia \cite{cuccagnageorgievvisciglia}  proved a decay estimate of the form \eqref{eq:infdec} for subcritical nonlinearities $\pm | \varphi_t|^{p-1} \varphi_t$ with $3<p<5$ and small initial data. The corresponding result for the critical nonlinearities $\pm | \varphi_t|^{p-1} \varphi_t$ with $p=3$ was proved by Germain, Pusateri and Rousset \cite{germainpusaterirousset}. Their proof relies on the use of the distorted Fourier transform and the analysis of an oscillatory integral.  A similar result was proved by Naumkin \cite{naumkinpotential,naumkinexceptional}. In dimension $d=3$, Pusateri and Soffer \cite{pusaterisoffer} considered the nonlinearity $-\varphi_t^2$ and small initial data. They showed that
\begin{equation}
	\| \varphi_t\|_\infty\le C(1+|t|)^{-(1+\alpha)}\ \textrm{for all } t\neq0
\end{equation}
for some $\alpha>0$. Two other results in dimension $d=3$ with external potentials, which we would like to mention, are scattering in $H^1$ for an external potential $V$ with small negative part by Hong \cite{hong}, and a classification of the dynamics of solutions to the cubic nonlinear Schrödinger equation with small initial data and a radial external potential $V$, which is such that the operator $-\Delta +V$ has exactly one negative eigenvalue, by Nakanishi \cite{nakanishi}.

\subsection{Strategy of the Proof}\label{sec:proof-strategy}

In order to compare $\gamma_{N,t}^{(1)}$ with {$|\varphi_t\rangle\langle\varphi_t|$}, we will need good control on the time evolution. Recall that $\varphi_t$ was the solution to the Hartree type equation \eqref{eq:Hartree}.
To this end, we will define a truncated dynamics in Section \ref{sec:truncation dynamics}, for which we have good bounds.
In Section \ref{sec:comparison-of-dynamics}, we then compare the dynamics with its truncated version, for which we deduce good bounds on the dynamics.

Moreover, we divide the dynamics into parity-conserving and non-parity-conserving parts.
In order to obtain these bounds, we estimate the time derivative and then use the Gr\"onwall lemma.
Since the estimate for the time derivative contains a factor of $\|\varphi_t\|_\infty$
and by Proposition \ref{prop:dispersive d>=3 small} {below}, we know from \eqref{eq:dispersive estimate for u main theorem} that $t \mapsto \|\varphi_t\|_\infty\in L^1(\bR)$, the bound we get after applying the Gr\"onwall lemma will be the time independent.

\subsection{Structure of the Paper}

In Section \ref{sec:preliminaries}, we review the necessary preliminaries. In particular, we review the result on the time decay estimate with an external potential, see \ref{prop:dispersive d>=3 small}, and the Fock Space formalism for the many-body problem. In Section \ref{sec:proof-main-theorem}, we prove Theorem \ref{thm:main} assuming Propositions \ref{prop:Et1} and \ref{prop:Et2}.
In Section \ref{sec:truncation dynamics}, we introduce the truncated dynamics and prove a bound for it.
Section \ref{sec:comparison-of-dynamics} is devoted to proving Propositions \ref{prop:Et1} and \ref{prop:Et2}. To this end, we show several results on comparison dynamics.

\subsection{Notation}
\begin{enumerate}
	\item The $L^p$-norm in $\bR^3$ can be denoted by $\|\cdot\|_{L^p(\bR^3)}$ or $\|\cdot\|_p$.
	\item Sometimes, when there can be no confusion, we omit the subscript for norms, for example,
	\[
	\|\cdot\|_{L^2(\bR^3)},\; \|\cdot\|_{\mathcal{F}},\;\text{or}\; \langle\cdot,\cdot\rangle_{\mathcal{F}}
	\]
	are just written as
	\[
	\|\cdot\|,\; \|\cdot\|,\;\text{or}\; \langle\cdot,\cdot\rangle,
	\]
	respectively.
	\item Our constant $C$ may change from line to line.
	\item The indicator function is denoted by $\chi$, for example
	\begin{equation*}
		\chi(N\le M)=\begin{cases}
			1 & \text{if }N\le M,\\
			0 & \text{if }N> M.
		\end{cases}
	\end{equation*}
	\item
	We define the Fourier transform of a function $f\in L^1(\bR^3)$ by
	\[
	\widehat{f}(\xi)=\int_{\bR^3}\mathrm{d}x\, e^{-2\pi \ii x \cdot \xi } f(x)\,.
	\]
\end{enumerate}

\noindent\textbf{Acknowledgements.}
The authors  would like to express their deepest gratitude to Phan Th\`anh Nam for his continued support and very helpful discussions. 
The authors also thank Arnaud Triay and Alessandro Michelangeli for helpful discussions. 
Partial support from the Deutsche For\-schungs\-ge\-mein\-schaft (DFG project Nr. 426365943) and Istituto Nazionale di Alta Matematica ``F. Severi,'' through the Intensive Period ``INdAM Quantum Meetings (IQM22)'' is acknowledged.

\section{Preliminaries}\label{sec:preliminaries}

\subsection{Time Decay Estimate under the Existence of an External Potential}
This subsection is devoted to reviewing the time decay estimate under the existence of external potential.

\begin{proposition}[{\cite[Theorem 1.1]{Dietze2021}} ]
\label{prop:dispersive d>=3 small}
	Let $d\ge 3$ and let $k\in\bN$ be the smallest even number with $k>\frac{d}{2}$. Let $V\in W^{k,\infty}(\bR^d)$ be a real-valued function and satisfy
	\begin{equation}\label{eq:dec}
		\|e^{-\ii t(-\Delta+V)}f\|_\infty\le C^V |t|^{-\frac{d}{2}}\|f\|_1
	\end{equation} 
	for every $f\in L^1(\bR^d)\cap L^2(\bR^d)$ and some constant $C^V\ge 1$. Let the interaction potential $w\in L^1(\bR^d)\cap L^{\frac{d}{2}}(\bR^d)$ be an even, real-valued function. Let $\varphi_0 \in H^k(\bR^d)$ and let $ \varphi_t \in C\left(\bR , H^2(\bR^d)\right)\cap C^1\left(\bR, H^{-1}(\bR^d)\right)$ be the unique global strong solution to the Hartree type equation 
	\begin{equation}\label{eq:nlsw}
		\begin{cases}
			\ii\partial_t  \varphi_t & =(-\Delta +V) \varphi_t+(w*| \varphi_t|^2) \varphi_t\\
			\varphi_t\big|_{t=0} & = \varphi_0.
		\end{cases}
	\end{equation}
	Assume that the initial data is sufficiently small, that is, 
	\begin{equation}\label{eq:initial data are small d>=3 small}
		\|e^{\ii (-\Delta+V)} \varphi_0\|_1\, , \| \varphi_0 \|_{H^k}\le\epsilon_0
	\end{equation}
	for some $\epsilon_0=\epsilon_0(d,\|V\|_{W^{k,\infty}}, C^V,\|w\|_1)>0$. 
	
	Then there exists a constant $C_0=C_0(d,\|V\|_{W^{k,\infty}}, C^V,\|w\|_1)\ge 1$ such that
	\begin{equation}\label{eq:dispersive estimate for u main theorem}
		\| \varphi_t\|_\infty\le\frac{C_0}{(1+|t|)^{\frac{d}{2}}}
	\end{equation}
	for all $t\ge 0$. Furthermore, if we assume that
	\begin{equation}
		\| e^{\ii (-\Delta+V)} ( \partial_t  \varphi_t)\big|_{t=0}\|_1\, ,\|(\partial_t \varphi_t)|_{t=0}\|_{H^k}\le\tilde\epsilon_0
	\end{equation}
	for some $\tilde\epsilon_0=\tilde \epsilon_0(d,\|V\|_{W^{k,\infty}}, C^V,\|w\|_1)>0$, then
	\begin{equation}
		\|\partial_t  \varphi_t\|_\infty\le  \frac{\tilde C_0}{(1+|t|)^{\frac{d}{2}}}
	\end{equation}
	for all $t\ge 0$, where $\tilde C_0=\tilde C_0(d,\|V\|_{W^{k,\infty}}, C^V,\|w\|_1)>0$. 
\end{proposition}
\vspace{2em}

\begin{remark}\phantom{ }
	\begin{enumerate}
		\item Let $V:\bR^3\to\bR$ and assume that 
		\begin{equation}\label{eq:c1}
			\int_{\bR^3\times \bR^3}\dx\dr{y} \frac{|V(x)\|V(y)|}{|x-y|^2}<(4\pi)^2
		\end{equation}
		and 
		\begin{equation}\label{eq:c2}
			\sup_{x\in\bR^3} \int_{\bR^3}\dr{y} \frac{|V(y)|}{|x-y|}<4\pi\, .
		\end{equation}
		Then \cite[Theorem 1.1]{rs} there exists a constant $C^V>0$ depending only on $V$ such that
		\begin{equation}
			\|e^{-\ri t{(-\Delta+V)}}f\|_\infty\le C^V|t|^{-\frac{3}{2}}\|f\|_1\ \textrm{for all } f\in L^1(\bR^3)\cap L^2(\bR^3)\,,\ t\in\bR\setminus\{0\} \, . 
		\end{equation}
		{ Hence, the combination of \eqref{eq:c1} and \eqref{eq:c2} is a sufficient condition for \eqref{eq:dec} in dimension three.}
		\item Proposition \ref{prop:dispersive d>=3 small} remains true for any global solution to the cubic non-linear Schrödinger equation: We may replace the interaction potential by $w=\pm\delta_0$, that is, the nonlinearity is given by $\pm|\varphi_t|^2 \varphi_t$.
	\end{enumerate}
\end{remark}

\begin{proof}[Idea of the proof]
	The proof uses ideas from Grillakis and Machedon \cite[Corollary 3.4]{GrillakisMachedon13} and Kato and Pusateri \cite{katopusateri}. We only describe the proof strategy for 
	\begin{equation*}
		\|\varphi_t\|_\infty\le\frac{C_0}{(1+|t|)^{\frac{d}{2}}}\ \text{for\ all\ } t\ge 0
	\end{equation*}
	as the proof of the corresponding estimate for $\|\partial_t \varphi_t\|_\infty$ is similar. Define
	\begin{equation*}
		M(T):= \sup_{0\le t\le T} (1+|t|)^{\frac{d}{2}}\|\varphi_t\|_\infty+\sup_{0\le t \le T}\|D^k \varphi_t\|_2+\|\varphi_0\|_2\, , 
	\end{equation*}
	and note that it suffices to show
	\begin{equation}\label{eq:goal}
		M(T)\le C_0\ \text{for\ every\ } T\ge 0
	\end{equation}
	for some $C_0>0$ to be determined. Using Duhamel's formula and the dispersive estimate 
	\begin{equation*}
		\|e^{-\ri t(-\Delta+V)}f\|_\infty\le C^V |t|^{-\frac{d}{2}}\|f\|_1\ \textrm{for all } f\in L^1(\bR^d)\cap L^2(\bR^d)
	\end{equation*} 
	for the operator $-\Delta+V$, which was part of our assumptions on the external potential $V$, we obtain after a computation
	\begin{equation*}
		M(T)\le\epsilon +CM(T)^3\, ,
	\end{equation*}
	where $\epsilon, C>0$ do not depend on $T$ and $\epsilon$ is small if the initial data is small. If $\epsilon> 0$ is small enough, then the graph of the function
	\begin{equation*}
		f:[0,\infty)\to\bR\,,\qquad f(x): =\epsilon +Cx^3-x
	\end{equation*}
	has two distinct non-negative zeros, see Figure \ref{fig:graph-of-f}
	
	\begin{figure}[ht]
		\centering
		\captionsetup{justification=centering,margin=2cm}
		\begin{tikzpicture}[scale=0.7]
			\begin{axis}[
				axis y line=center,
				axis x line=middle, 
				axis on top=true,
				xmin=-0.1,
				xmax=0.6,
				ymin=-0.1,
				ymax=0.6,
				height=10.0cm,
				width=10.0cm,
				xtick={0.1,0.2,0.3,0.4,0.5},
				ytick={0.1,0.2,0.3,0.4,0.5},
				]
				\node[below left] at (0,0) {$0$};
				\addplot [domain=-0.1:0.5, samples=50, mark=none, ultra thick, blue] {0.1 + 8*(x)^3- (x)};
				\node [left, blue] at (axis cs: 0.38,0.5) {\large $f(x)=\epsilon+C x^3-x$};
				\fill[red] (axis cs: 0.110916,0) circle [radius=3pt] node[above right] {\large $C_0$};
			\end{axis}
		\end{tikzpicture}
		\caption{Graph of $f$}
		\label{fig:graph-of-f}
	\end{figure}
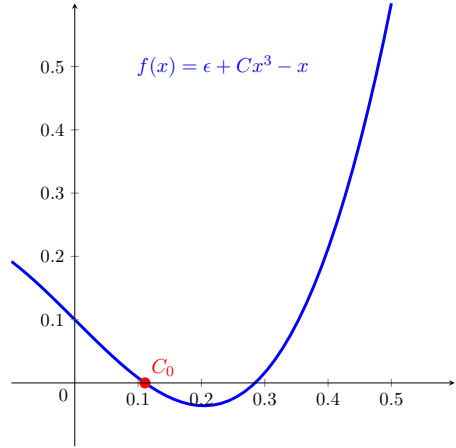
	
	Denote by $C_0$ the first zero of $f$. By a blow-up criterion for $H^k$-solutions similar to \cite[Theorem 4.10.1]{cazenave}, there exists $T_{\max}\in (0,\infty]$ such that $M\in C\left( [0, T_{\max})\right)$. Furthermore, if $T_{\max}<\infty$, then $\lim_{T\uparrow T_{\max}}M(T)=\infty$. If $M(0)\le C_0$, then by the smallness of the initial data, we deduce that $T_{\max}=\infty$ and {\eqref{eq:goal} holds.}
	\end{proof}
	
	\vspace{2em}
	
	\begin{remark}
		By writing $ \widetilde{u}= u/\alpha$ for some $\alpha> 0$ to be chosen, \eqref{eq:nlsw} becomes
		\begin{equation}
			\begin{cases}
				\ii\partial_t \widetilde{u}_t & =(-\Delta +V) \widetilde{u}_t+\alpha^2(w*|\widetilde{u}_t|^2)\widetilde{u}_t\\
				\widetilde{u}_t\big|_{t=0} & =\frac 1 \alpha \varphi_0 =:\widetilde{u}_0\,.
			\end{cases}
		\end{equation}
		If $ \alpha $ is small, this is an equation with weak coupling for large initial data. Hence, the results in Proposition \ref{prop:dispersive d>=3 small} can be converted to the corresponding results for weakly interacting Hartree/NLS equations with large initial data.
	\end{remark}
	
	\subsection{Fock Space}
	To investigate the dynamics of a Bose gas with $N$ particles, we embed our problem into a bigger space, the so-called Fock space. We briefly review the well-known Fock space formalism. The bosonic Fock space is a Hilbert space given by
	\[
	\cF=\bigoplus_{n\geq0}L^{2}(\bR^{3})^{\otimes_{s}n},
	\]
	where $L^{2}(\bR^{3})^{\otimes_{s}n}$ is the symmetric subspace of $L^2(\bR^{3n})$, that is, such that, for 
	\[
	g(x_1,\dots,x_n) = g(x_{\sigma(1)},\dots,x_{\sigma(n)})
	\]
	for all $g(x_1,\dots,x_n)\in L^{2}(\bR^{3})^{\otimes_{s}n}$ and all $\sigma\in S_n$, where $S_n$ denotes the set of all permutations of $\{ 1,\ldots,n\}$.
	We denote an element (or state) $\psi\in\cF$ by
	\[
	\psi = \psi^{(0)} \oplus \psi^{(1)} \oplus \psi^{(2)} \oplus \dots =\left(\psi^{(n)}\right)_{n\ge0}
	\]
	where $L^{2}(\bR^{3})^{\otimes_{s}n}$ for all $n\ge0$.
	The inner product on $\cF$ is defined by
	\begin{equation}
		\begin{aligned}
			\langle \psi_{1},\psi_{2}\rangle
			&:=\sum_{n\geq0}\langle \psi_{1}^{(n)},\psi_{2}^{(n)}\rangle_{L^{2}(\bR^{3n})}\\
			&=\overline{ \psi_{1}^{(0)} } \psi_{2}^{(0)}+\sum_{n\geq 1}\int_{\bR^{3n}} \dx _{1}{{\dots}} \dx _{n}\,\overline{\psi_{1}^{(n)}(x_{1},{{\dots}},x_{n})}\psi_{2}^{(n)}(x_{1},{{\dots}},x_{n}).
		\end{aligned}
	\end{equation}
	For $f\in L^2(\bR^3)$, the creation operator $a^{*}(f)$ and the annihilation operator $a(f)$ on $\cF$ are defined by
	\begin{align}
		(a^{*}(f)\psi)^{(n)}(x_{1},{{\dots}},x_{n})&=\frac{1}{\sqrt{n}}\sum_{j=1}^{n}f(x_{j})\psi^{(n-1)}(x_{1},{{\dots}},x_{j-1},x_{j+1},{{\dots}},x_{n}) \label{eq:creation}\\
		\intertext{and}
		(a(f)\psi)^{(n)}(x_{1},{{\dots}},x_{n})&=\sqrt{n+1}\int_{\bR^{3}} \dr{x} \overline{f(x)}\psi^{(n+1)}(x,x_{1},{{\dots}},x_{n}).\label{eq:annihilation}
	\end{align}
	Define
	\begin{equation}\label{eq:dephi}
		\phi(f):=a^{*}(f)+a(f).
	\end{equation}
	Note that $\phi(f)$ is self-adjoint, while the creation operator $a^{*}(f)$ and the annihilation operator $a(f)$ are not self-adjoint.
	We also use operator-valued distributions $a_{x}^{*}$ and $a_{x}$ satisfying
	\begin{equation}\label{eq:deancr}
		a^{*}(f)=\int_{\bR^{3}} \dx \,f(x)a_{x}^{*}, \qquad a(f)=\int_{\bR^{3}} \dx \,\overline{f(x)}a_{x}
	\end{equation}
	for any $f\in L^{2}(\bR^{3})$. The annihilation and creation operator satisfy the canonical commutation relation (CCR), which are given by
	\[
	[a(f),a^{*}(g)]=\langle f,g\rangle_{L^{2}(\bR^{3})},\quad[a(f),a(g)]=[a^{*}(f),a^{*}(g)]=0,
	\]
	and
	\[
	[a_{x},a_{y}^{*}]=\delta(x-y),\quad[a_{x},a_{y}]=[a_{x}^{*},a_{y}^{*}]=0.
	\]
	The number operator $\cN $ is defined by
	\begin{equation} \label{eq:number operator}
		\cN :=\int_{\bR^{3}}  \dx\, a_{x}^{*}a_{x}
	\end{equation}
	and it satisfies that $(\cN \psi)^{(n)}=n\psi^{(n)}$. The domain of the number operator $\cN $ is given by
	\begin{equation} 
		D(\cN) = \left\{ \psi \in \mathcal{F} : \sum_{n \geq 1} n^{2} \| \psi^{(n)} \|^2 < \infty \right\}
	\end{equation}
	
	For any bounded operator $J:L^{2}(\bR^{3})\to L^{2}(\bR^{3})$,
	its second quantization $d\Gamma(J):D(\cN)\to\cF$ is given by
	\[
	(d\Gamma(J)\psi)^{(n)}=\sum_{j=1}^{n}J_{j}\psi^{(n)}
	\]
	where $J_{j}=1\otimes{{\dots}}\otimes J\otimes{{\dots}}\otimes1$ is the operator $J$ acting only on the $j$-th variable. For any compact operator $J:L^{2}(\bR^{3})\to L^{2}(\bR^{3})$ with kernel $ J (x; y)$, we can write its second quantization $d\Gamma(J)$ in terms of operator-valued distributions as
	\begin{equation}\label{eq:dgjop}
		d\Gamma(J) = \int_{\bR^3 \times \bR^3}\dx\dy\,
		J(x;y)a_{x}^{*}a_{y}.
	\end{equation}
	
	Intuitively speaking, both creation and annihilation operators behave like $\cN^{1/2}$ as the following lemma shows:
	\begin{lemma}[{\cite[Lemma 2.1]{RodnianskiSchlein2009}} ]
		For $\alpha >0$, let $D(\cN^{\alpha}) = \{ \psi \in \mathcal{F} : \sum_{n \geq 1} n^{2\alpha} \| \psi^{(n)} \|^2 < \infty \}$ denote the domain of the operator $\cN^{\alpha}$. 
		For any $f \in L^2 (\bR^3)$ and any $\psi \in D (\cN^{1/2})$, we have
		\begin{equation}\label{eq:bd-a}
			\begin{split}  
				\| a(f) \psi \| & \leq \| f \| \, \| \cN^{1/2} \psi \|, \\
				\| a^* (f) \psi \| &\leq \| f \| \, \| (\cN+1)^{1/2} \psi \|, \\
				\| \phi (f) \psi \| &\leq 2 \| f \| \| ( \cN+ 1 )^{1/2} \psi \| \, . 
			\end{split} 
		\end{equation}
		Moreover, for any bounded one-particle operator $J$ on $L^2 (\bR^3)$ and for every $\psi \in D (\cN )$, we find
		\begin{equation}
			\| d\Gamma (J) \psi \| \leq \| J \| \| \cN  \psi \|  \, .\label{eq:J-bd} 
		\end{equation}
	\end{lemma}
	
	For $f\in L^{2}(\bR^{3})$, the Weyl operator $W(f)$ is defined by
	\[
	W(f):=\exp(a^{*}(f)-a(f)),
	\]
	and it satisfies
	\[
	W(f)=e^{-\Vert f\Vert^{2}/2}\exp(a^{*}(f))\exp(-a(f)).
	\]
	Coherent states $\psi(f)$ can be defined in terms of the Weyl operator as
	\begin{equation} \label{Weyl_f}
		\psi(f):=W(f)\Omega = e^{-\Vert f\Vert^{2}/2}\exp(a^{*}(f))\Omega =e^{-\Vert f\Vert^{2}/2}\sum_{n\geq0}\frac{1}{\sqrt{n!}}f^{\otimes n}.
	\end{equation}
	
	The following lemmata indicate useful properties of the Weyl operator and coherent states:
	\begin{lemma}[{\cite[Lemma 2.2]{RodnianskiSchlein2009}} ] 
	\label{lem:Basic-Weyl}
		Let $f,g\in L^{2}(\bR^{3})$.
		\begin{enumerate}
			\item The commutation relation between the Weyl operators is given by
			\[
			W(f)W(g)=W(g)W(f)e^{-2\ii  \cdot \ImPart \langle f,g\rangle}=W(f+g)e^{-\ii  \cdot \ImPart \langle f,g\rangle}.
			\]
			
			\item The Weyl operator is unitary and satisfies
			\[
			W(f)^{*}=W(f)^{-1}=W(-f).
			\]
			
			\item The coherent states are eigenvectors of annihilation operators, i.e.,
			\[
			a_{x}\psi(f)=f(x)\psi(f)\quad\rightarrow\quad a(g)\psi(f)=\langle g,f\rangle_{L^{2}}\psi(f).
			\]
			\item The commutation relation between the Weyl operator and the annihilation operator (or the creation operator) is
			\[
			W^{*}(f)a_{x}W(f)=a_{x}+f(x)\quad\text{and}\quad W^{*}(f)a_{x}^{*}W(f)=a_{x}^{*}+\overline{f(x)}.
			\]
			
			\item The distribution of $\cN $ with respect to the coherent state $\psi(f)$ is Poisson. In particular,
			\[
			\langle \psi(f),\cN \psi(f)\rangle=\Vert f\Vert^{2}, \qquad \langle \psi(f),\cN^{2}\psi(f)\rangle-\langle \psi(f),\cN \psi(f)\rangle ^{2}=\Vert f\Vert^{2}.
			\]
		\end{enumerate}
	\end{lemma}
	
	Let
	\begin{equation} \label{eq:d_N}
		d_{N}:=\frac{\sqrt{N!}}{N^{N/2}e^{-N/2}}.
	\end{equation}
	Note that $C^{-1} N^{1/4} \leq d_N \leq C N^{1/4}$ for some constant $C>0$ independent of $N$, which can be easily checked by using Stirling's formula.
	\begin{lemma}[{\cite[Lemma 6.3]{ChenLee2011}} ]\label{lem:coherent_all}
		There exists a constant $C>0$ independent of $N$ such that, for any $\varphi_0\in L^{2}(\bR^{3})$ with $\|\varphi_0\|=1$, we have
		\[
		\left\Vert (\cN+1)^{-1/2}W^{*}(\sqrt{N}\varphi_0)\frac{(a^{*}(\varphi_0))^{N}}{\sqrt{N!}}\Omega\right\Vert \leq\frac{C}{d_{N}}.
		\]
	\end{lemma}

	\begin{lemma}[{\cite[Lemma 7.2]{Lee2013}} ]\label{lem:coherent_even_odd}
		Let $P_m$ be the projection onto the $m$-particle sector of the Fock space $\mathcal{F}$ for a non-negative integer $m$. Then, for any non-negative integer $k\leq(1/2)N^{1/3}$ and any $\varphi_0\in L^{2}(\bR^{3})$ with $\|\varphi_0\|=1$,
		\[
		\left\Vert P_{2k}W^{*}(\sqrt{N}\varphi_0)\frac{(a^{*}(\varphi_0))^{N}}{\sqrt{N!}}\Omega\right\Vert \leq\frac{2}{d_{N}}
		\]
		and
		\[
		\left\Vert P_{2k+1}W^{*}(\sqrt{N}\varphi_0)\frac{(a^{*}(\varphi_0))^{N}}{\sqrt{N!}}\Omega\right\Vert \leq\frac{2(k+1)^{3/2}}{d_{N}\sqrt{N}}.
		\]
	\end{lemma}
	
	\section{Proof of the Main Result}\label{sec:proof-main-theorem}
	
	To consider the problem in the Fock space formalism, we extend the Hamiltonian $H_{N}$ in \eqref{eq:N_body_Hamiltonian} to the Fock space by
	\begin{equation} \label{eq:Fock_space_Hamiltonian}
		\cH _{N}:=\int_{\bR^3} \dx \,a_{x}^{*}\big(-\Delta_{x}+V(x)\big)a_{x}+\frac{\lambda}{2N}\int_{\bR^3 \times \bR^3} \dx  \dy \,w(x-y)a_{x}^{*}a_{y}^{*}a_{y}a_{x}.
	\end{equation}
	With this definition, we have $(\cH _{N}\psi)^{(N)}=H_{N}\psi^{(N)}$ for any $\psi\in\mathcal{F}$.
	The kernel of the one-particle marginal density $\gamma_{\psi}^{(1)}$ associated with $\psi$ is
	\begin{equation} \label{eq:Kernel_gamma}
		\gamma_{\psi}^{(1)}(x;y)=\frac{1}{\langle \psi,\cN \psi\rangle}\langle \psi,a_{y}^{*}a_{x}\psi\rangle .
	\end{equation}
	Note that $\gamma_{\psi}^{(1)}$ is a trace class operator on $L^{2}(\bR^{3})$ and $\text{Tr }\gamma_{\psi}^{(1)}=1$. {Definition} \eqref{eq:Kernel_gamma} is equivalent to \eqref{eq:Kernel_of_Marginal_Density} in the sense that $\gamma_{\psi}^{(1)} = \gamma_{N,t}^{(1)}$ for every $\psi =\left(1_{n=N}\psi_{N,t}\right)_{n\ge0}\in\cF $ and $\psi_{N,t}\in L^{2}(\bR^{3})^{\otimes_{s}N}$.
	
	Let $\gamma_{N,t}^{(1)}(x;y)$ for $x,y\in\bR^3$ be the kernel of the one-particle marginal density associated with the time evolution of the factorized state $\varphi_0^{\otimes N}$ with respect to the Hamiltonian $\mathcal{H}_N$. By the equivalence of \eqref{eq:Kernel_gamma} and \eqref{eq:Kernel_of_Marginal_Density},
	\begin{equation} \label{eq:marginal_factorized}
		\begin{split}
			\gamma_{N,t}^{(1)}(x;y) & =\frac{\langle e^{-\ii \mathcal{H}_{N}t}\varphi_0^{\otimes N},a_{y}^{*}a_{x}e^{-\ii \mathcal{H}_{N}t}\varphi_0^{\otimes N}\rangle}{\langle e^{-\ii \mathcal{H}_{N}t}\varphi_0^{\otimes N},\cN  e^{-\ii \mathcal{H}_{N}t}\varphi_0^{\otimes N}\rangle} =\frac{1}{N}\left\langle \varphi_0^{\otimes N}, e^{\ii \mathcal{H}_{N}t} a_{y}^{*}a_{x}e^{-\ii \mathcal{H}_{N}t}\varphi_0^{\otimes N}\right\rangle\\
			& =\frac{1}{N}\left\langle \frac{\left(a^{*}(\varphi_0)\right)^{N}}{\sqrt{N!}}\Omega,e^{\ii \mathcal{H}_{N}t}a_{y}^{*}a_{x}e^{-\ii \mathcal{H}_{N}t}\frac{\left(a^{*}(\varphi_0)\right)^{N}}{\sqrt{N!}}\Omega\right\rangle .
		\end{split}
	\end{equation}
	{Let us for a moment discuss the time evolution of a coherent state, taken as $W^{*}(\sqrt{N}\varphi_{s})\Omega$. We expand}
	$a_{y}^{*}a_{x}$ around $N \overline{\varphi_t(y)} \varphi_t(x)$, then we are lead to consider the operator
	\begin{equation} \label{eq:introducing U}
		\begin{split}
			&W^{*}(\sqrt{N}\varphi_{s}) e^{\ii \mathcal{H}_{N}(t-s)} (a_x - \sqrt{N} \varphi_t(x) ) e^{-\ii \mathcal{H}_{N}(t-s)} W(\sqrt{N}\varphi_{s}) \\
			&= W^{*}(\sqrt{N}\varphi_{s}) e^{\ii \mathcal{H}_{N}(t-s)} W(\sqrt{N}\varphi_{t}) a_x W^{*}(\sqrt{N}\varphi_{t}) e^{-\ii \mathcal{H}_{N}(t-s)} W(\sqrt{N}\varphi_{s}).
		\end{split}
	\end{equation}
	After a computation, we obtain
	\begin{equation} \label{eq:derivative decomposition}
		\ii  \partial_t \, W^{*}(\sqrt{N}\varphi_{t}) e^{-\ii \mathcal{H}_{N}(t-s)} W(\sqrt{N}\varphi_{s}) =: \Big( \sum_{k=0}^{4} \cL_k (t;s) \Big) W^{*}(\sqrt{N}\varphi_{t}) e^{-\ii \mathcal{H}_{N}(t-s)} W(\sqrt{N}\varphi_{s}),
	\end{equation}
	where $\cL_k$ is an operator with $k$ creation and/or annihilation operators. The exact formulas for $\cL_k$ are as follows:
	\begin{align}
		{{\cL_0 (t;s)}} &:= \frac{N\lambda}{2} \int_s^t \mathrm{d}\tau \int_{\bR^3} \dx\, (w *| \varphi_{\tau}|^2 ) (x) |\varphi_{\tau} (x)|^2, \\
		\cL_1 (t;s)& :=\cL_1  = 0, \\
		\cL_2 (t;s)& := \cL_{2}(t) = \int_{\bR^3} \dx\, a_{x}^{*}\big(-\Delta_x+V(x)\big)a_{x}+\lambda\int_{\bR^3} \dx\,(w*|\varphi_{t}|^{2})(x)a_{x}^{*}a_{x}\notag \\
		&\qquad+\lambda\int_{\bR^3\times\bR^3}\dx\dy\, w (x-y)\overline{\varphi_{t}(x)}\varphi_{t}(y)a_{y}^{*}a_{x} \notag\\
		& \qquad+\frac{\lambda}{2}\int_{\bR^3 \times \bR^3} \dx\dy\,  w(x-y)\big(\varphi_{t}(x)\varphi_{t}(y)a_{x}^{*}a_{y}^{*}+\overline{\varphi_{t}(x)}\,\overline{\varphi_{t}(y)}a_{x}a_{y}\big), \label{eq:L_2} \\
		\cL_3 (t;s)& := \cL_{3}(t) = \frac{\lambda}{\sqrt{N}}\int_{\bR^3 \times \bR^3} \dx\dy\,  w(x-y)\big(\varphi_{t}(y)a_{x}^{*}a_{y}^{*}+\overline{\varphi_{t}(y)}a_{x}^{*}a_{y}\big)a_{x}, \label{eq:L_3} \\
		\cL_4 (t;s)& := \cL_{4} = \frac{\lambda}{2N}\int_{\bR^3 \times \bR^3} \dx\dy\, w(x-y)a_{x}^{*}a_{y}^{*}a_{x}a_{y}. \label{eq:L_4}
	\end{align}
	
	Defining the unitary operator $\cU(t;s)$ by
	\begin{equation}\label{eq:deuts}
		\cU (t;s) := e^{\ii\, {\omega(t;s)}} W^{*}(\sqrt{N}\varphi_{t}) e^{-\ii \mathcal{H}_{N}(t-s)} W(\sqrt{N}\varphi_{s}),
	\end{equation}
	with the phase factor
	\[
	\omega(t;s) = \int_s^t \dr{\tau} {\cL_0 (\tau;s)},
	\]
	we get
	\begin{equation}\label{eq:def_mathcalU}
		\ii \partial_{t}\,\cU(t;s)=(\cL_{2}+\cL_{3}+\cL_{4})\,\cU(t;s)=:\cL_N(t)\,\cU(t;s)\quad\text{and}\quad\cU(s;s)=I
	\end{equation}
	and
	\begin{equation}
		W^{*}(\sqrt{N}\varphi_{s})e^{\ii \mathcal{H}_{N}(t-s)}\big(a_{x}-\sqrt{N}\varphi_{t}(x)\big)e^{-\ii \mathcal{H}_{N}(t-s)}W(\sqrt{N}\varphi_{s})=\cU^{*}(t;s)\,a_{x}\,\cU(t;s),
	\end{equation}
	where we used \eqref{eq:introducing U} and the fact that the operator $e^{\ii\, {\omega(t;s)}}$ is just a multiplication by a complex number. 
	
	We let $\widehat{\cL}:=\cL_{2}+\cL_{4}$
	and define the unitary operator $\widehat{\cU}(t;s)$ by
	\begin{equation}\label{eq:def_mathcalwidehatU}
		\ii \partial_{t}\,\widehat{\cU}(t;s)
		=
		\widehat{\cL}(t)\,\widehat{\cU}(t;s)\quad\text{and} \quad\widehat{\cU}(s;s)=1.
	\end{equation}
	Since $\widehat{\cL}$ does not change the parity of the number of particles,
	\begin{equation}
		\langle \Omega,\widehat{\cU}^{*}(t;0)\,a_{y}\,\widehat{\cU}(t;0)\Omega\rangle=\langle \Omega,\widehat{\cU}^{*}(t;0)\,a_{x}^{*}\,\widehat{\cU}(t;0)\Omega\rangle=0.
		\label{eq:Parity_Consevation_hat}
	\end{equation}
	
	Now, we have the following bounds for $E_{t}^{(1)}(J)$ and $E_{t}^{(2)}(J)$, which will be defined and Proposition \ref{prop:Et1} and Proposition \ref{prop:Et2}. {They will be employed in the proof of Theorem \ref{thm:main}.}
	\begin{proposition}
		\label{prop:Et1}Suppose that the assumptions in Theorem \ref{thm:main}
		hold. For any compact Hermitian operator $J$ on $L^{2}(\bR^{3})$, let
		\[
		E_{t}^{(1)}(J):=\frac{d_{N}}{N}\left\langle W^{*}(\sqrt{N}\varphi_0)\frac{(a^{*}(\varphi_0))^{N}}{\sqrt{N!}}\Omega,\cU^{*}(t;0)d\Gamma(J)\cU(t;0)\Omega\right\rangle.
		\]
		Then, there exist constants $C$ depending only on 
		$\|w\|_{L^{1}(\bR^{3})}$, $\|w\|_{L^{2}(\bR^{3})}$, $\|V\|_{W^{2,\infty}}$ and $C^V$ such that
		\[
		|E_{t}^{(1)}(J)|\leq\frac{C\|J\|}{N}.
		\]
	\end{proposition}
	
	\begin{proposition}
		\label{prop:Et2}Suppose that the assumptions in Theorem \ref{thm:main}
		hold. For any compact Hermitian operator $J$ on $L^{2}(\bR^{3})$, let
		\[
		E_{t}^{(2)}(J):=\frac{d_{N}}{\sqrt{N}}\left\langle W^{*}(\sqrt{N}\varphi_0)\frac{(a^{*}(\varphi_0))^{N}}{\sqrt{N!}}\Omega,\cU^{*}(t;0)\phi(J\varphi_{t})\cU(t;0)\Omega\right\rangle.
		\]
		Then, there exist constants $C$ depending only on 
		$\|w\|_{L^{1}(\bR^{3})}$, $\|w\|_{L^{2}(\bR^{3})}$, $\|V\|_{W^{2,\infty}}$ and $C^V$ such that
		\[
		|E_{t}^{(2)}(J)|\leq C\|J\|N^{-1}.
		\]
	\end{proposition}
	The proofs of these propositions will be given in Section \ref{sec:comparison-of-dynamics}.
	
	\begin{proof}[Proof of Theorem~\ref{thm:main}]		
		Recall that
		\begin{equation}
			\gamma_{N,t}^{(1)}(x;y) =\frac{1}{N}\left\langle \frac{\left(a^{*}(\varphi_0)\right)^{N}}{\sqrt{N!}}\Omega,e^{i\mathcal{H}_{N}t}a_{y}^{*}a_{x}e^{-i\mathcal{H}_{N}t}\frac{\left(a^{*}(\varphi_0)\right)^{N}}{\sqrt{N!}}\Omega\right\rangle.
		\end{equation}
		From the definition of the creation operator in \eqref{eq:creation} and the definition of $ d_N$ in \eqref{eq:d_N}, we can easily find that
		\begin{equation} \label{eq:coherent_vec}
			\{0,0,{{\dots}},0,\varphi_0^{\otimes N},0,{{\dots}}\}=\frac{\left(a^{*}(\varphi_0)\right)^{N}}{\sqrt{N!}}\Omega,
		\end{equation}
		where the $\varphi_0^{\otimes N}$ on the left-hand side is in the $N$-th sector of the Fock space. Recall that $P_N$ is the projection onto the $N$-particle sector of the Fock space. From \eqref{Weyl_f}, we find that
		\[
		\frac{\left(a^{*}(\varphi_0)\right)^{N}}{\sqrt{N!}}\Omega=\frac{\sqrt{N!}}{N^{N/2}e^{-N/2}}P_{N}W(\sqrt{N}\varphi_0)\Omega=d_{N}P_{N}W(\sqrt{N}\varphi_0)\Omega.
		\]
		Since $\mathcal{H}_N$ does not change the number of particles, we also have that
		\begin{align*}
			\gamma_{N,t}^{(1)}(x;y) & =\frac{1}{N}\left\langle \frac{\left(a^{*}(\varphi_0)\right)^{N}}{\sqrt{N!}}\Omega,e^{\mathrm{i}\mathcal{H}_{N}t}a_{y}^{*}a_{x}e^{-\mathrm{i}\mathcal{H}_{N}t}\frac{\left(a^{*}(\varphi_0)\right)^{N}}{\sqrt{N!}}\Omega\right\rangle \\
			& =\frac{d_{N}}{N}\left\langle \frac{\left(a^{*}(\varphi_0)\right)^{N}}{\sqrt{N!}}\Omega,e^{\mathrm{i}\mathcal{H}_{N}t}a_{y}^{*}a_{x}e^{-\mathrm{i}\mathcal{H}_{N}t}P_{N}W(\sqrt{N}\varphi_0)\Omega\right\rangle \\
			& =\frac{d_{N}}{N}\left\langle \frac{\left(a^{*}(\varphi_0)\right)^{N}}{\sqrt{N!}}\Omega,P_{N}e^{\mathrm{i}\mathcal{H}_{N}t}a_{y}^{*}a_{x}e^{-\mathrm{i}\mathcal{H}_{N}t}W(\sqrt{N}\varphi_0)\Omega\right\rangle \\
			& =\frac{d_{N}}{N}\left\langle \frac{\left(a^{*}(\varphi_0)\right)^{N}}{\sqrt{N!}}\Omega,e^{\mathrm{i}\mathcal{H}_{N}t}a_{y}^{*}a_{x}e^{-\mathrm{i}\mathcal{H}_{N}t}W(\sqrt{N}\varphi_0)\Omega\right\rangle, \\
		\end{align*}
		where we used that $P_{N}\frac{\left(a^{*}(\varphi_0)\right)^{N}}{\sqrt{N!}}\Omega =\frac{\left(a^{*}(\varphi_0)\right)^{N}}{\sqrt{N!}}\Omega$ in the last step.
		To simplify it further, we use the relation
		\begin{equation}\label{eq:eae}
			e^{\mathrm{i}\mathcal{H}_{N}t}a_{x}e^{-\mathrm{i}\mathcal{H}_{N}t}=W(\sqrt{N}\varphi_0)\cU^{*}(t;0)(a_{x}+\sqrt{N}\varphi_{t}(x))\cU(t;0)W^{*}(\sqrt{N}\varphi_0),
		\end{equation}
		which follows from the first equality in Lemma \ref{lem:Basic-Weyl}(4), the definition of $\cU$ in \eqref{eq:deuts} 
		and the unitary of the Weyl operator, see Lemma \ref{lem:Basic-Weyl}(2). By \eqref{eq:eae}
		and an analogous result for the creation operator, we obtain that
		\begin{align*}
			\gamma_{N,t}^{(1)}(x;y) & =\frac{d_{N}}{N}\left\langle \frac{\left(a^{*}(\varphi_0)\right)^{N}}{\sqrt{N!}}\Omega,e^{\mathrm{i}\mathcal{H}_{N}t}a_{y}^{*}a_{x}e^{-\mathrm{i}\mathcal{H}_{N}t}W(\sqrt{N}\varphi_0)\Omega\right\rangle \\
			& =\frac{d_{N}}{N}\left\langle \frac{\left(a^{*}(\varphi_0)\right)^{N}}{\sqrt{N!}}\Omega,W(\sqrt{N}\varphi_0)\cU^{*}(t;0)\left(a_{y}^{*}+\sqrt{N}\,\overline{\varphi_{t}\left(y\right)}\right)\left(a_{x}+\sqrt{N}\varphi_{t}(x)\right)\cU(t;0)\Omega\right\rangle.
		\end{align*}

		Thus,
		\begin{align}
			\begin{split}\label{eq:gaphyphx}
				\gamma_{N,t}^{(1)}(x;y)-\overline{\varphi_{t}\left(y\right)}\varphi_t(x) & =\frac{d_{N}}{N}\left\langle \frac{\left(a^{*}(\varphi_0)\right)^{N}}{\sqrt{N!}}\Omega,W(\sqrt{N}\varphi_0)\cU^{*}(t;0)a_{y}^{*}a_{x}\cU(t;0)\Omega\right\rangle \\
				& \quad+\overline{\varphi_{t}(y)}\frac{d_{N}}{\sqrt{N}}\left\langle \frac{\left(a^{*}(\varphi_0)\right)^{N}}{\sqrt{N!}}\Omega,W(\sqrt{N}\varphi_0)\cU^{*}(t;0)a_{x}\cU(t;0)\Omega\right\rangle \\
				& \quad+\varphi_{t}(x)\frac{d_{N}}{\sqrt{N}}\left\langle \frac{\left(a^{*}(\varphi_0)\right)^{N}}{\sqrt{N!}}\Omega,W(\sqrt{N}\varphi_0)\cU^{*}(t;0)a_{y}^{*}\cU(t;0)\Omega\right\rangle.
			\end{split}
		\end{align}
		Recall the definition of $E_{t}^{(1)}(J)$ and $E_{t}^{(2)}(J)$ in Propositions \ref{prop:Et1} and \ref{prop:Et2}. For any compact one-particle Hermitian operator $J$ on $L^{2}(\mathbb{R}^{3})$, we have
		\begin{align*}
			\Tr\left( J\left({\gamma}_{N,t}^{(1)}-|\varphi_{t}\rangle \langle \varphi_{t}|\right)\right) & =\int_{\bR^3 \times \bR^3}\dx\dy\,
			J(x;y)\left(\gamma_{N,t}^{(1)}(y;x)-\varphi_{t}(y)\overline{\varphi_{t}\left(x\right)}\right)\\
			& =\frac{d_{N}}{N}\left\langle \frac{\left(a^{*}(\varphi_0)\right)^{N}}{\sqrt{N!}}\Omega,W(\sqrt{N}\varphi_0)\cU^{*}(t;0)d\Gamma(J)\cU(t;0)\Omega\right\rangle \\
			& \quad+\frac{d_{N}}{\sqrt{N}}\left\langle \frac{\left(a^{*}(\varphi_0)\right)^{N}}{\sqrt{N!}}\Omega,W(\sqrt{N}\varphi_0)\cU^{*}(t;0)\phi(J\varphi_{t})\cU(t;0)\Omega\right\rangle \\
			& =E_{t}^{(1)}(J)+E_{t}^{(2)}(J).
		\end{align*}
		The second step can be seen by using the expression for $d\Gamma(J)$ in terms of operator-valued distributions in \eqref{eq:dgjop}, \eqref{eq:gaphyphx}, the definition of annihilation and creation operators in terms of operator-valued distributions in \eqref{eq:deancr} and the definition of $\phi$ in \eqref{eq:dephi}. Thus, from Propositions \ref{prop:Et1} and \ref{prop:Et2}, we find that
		\begin{equation}\label{eq:trjpro3132}
			\left|\Tr\left( J\big({\gamma}_{N,t}^{(1)}-|\varphi_{t}\rangle \langle \varphi_{t}|\big)\right)\right|\leq C\|J\|N^{-1} 
		\end{equation}
		Since the space of compact operators is the dual to the space of the trace class operators, and since ${\gamma}_{N,t}^{(1)}$ and $|\varphi_{t}\rangle \langle \varphi_{t}|$
		are Hermitian and trace class, there exists an orthonormal basis $\left( f_k\right)_{k\in\bN}\subset L^{2}(\bR^{3})$ and a sequence of real numbers $\left( s_k\right)_{k\in\bN}\subset \bR$ with $\sum_{k\in\bN}s_k<\infty$ such that
		\begin{equation}
			{\gamma}_{N,t}^{(1)}-|\varphi_{t}\rangle \langle \varphi_{t}|=\sum_{k\in\bN}s_k|f_k\rangle \langle f_k|.
		\end{equation}
		Now for any $K\in\bN$ choose the Hermitian  compact operator
		\begin{equation}
			J_K:=\sum_{k=1}^K \sign\left(s_k\right)|f_k\rangle \langle f_k|,
		\end{equation}
		and note that $\left\Vert J_K\right\Vert\le1$. We have
		\begin{equation}
			\Tr\left|{\gamma}_{N,t}^{(1)}-|\varphi_{t}\rangle \langle \varphi_{t}|\right|=\sum_{k\in\bN}s_k=\lim_{K\to\infty}\sum_{k=1}^Ks_k =\lim_{K\to\infty}\left|\Tr\left( J_K\big({\gamma}_{N,t}^{(1)}-|\varphi_{t}\rangle \langle \varphi_{t}|\big)\right)\right|.
		\end{equation}
		Combining this with \eqref{eq:trjpro3132}, we obtain
		\[
		\Tr\left|{\gamma}_{N,t}^{(1)}-|\varphi_{t}\rangle \langle \varphi_{t}|\right|\leq
			C N^{-1}.
		\]
		which shows the theorem.
	\end{proof}
	
	\section{Truncation Dynamics}\label{sec:truncation dynamics}
	
	First, we introduce a truncated time-dependent generator with fixed $M>0$ as follows:
	\begin{align*}
		\mathcal{L}_{N}^{(M)}(t) & =\int_{\bR^3}\dx\,a_{x}^{*}(-\Delta_{x}+V(x))a_{x}\\
		&\quad+\lambda\int_{\bR^3}\dx\,\left(w*|\varphi_{t}|^{2}\right)(x)a_{x}^{*}a_{x}+\lambda \int_{\bR^3\times\bR^3}\dx\dy\, w(x-y)\overline{\varphi_{t}(x)}\varphi_{t}(y)a_{y}^{*}a_{x}\\
		& \quad+\frac{\lambda}{2}\int_{\bR^3\times\bR^3}\dx\dy\,w(x-y)\left(\varphi_{t}(x)\varphi_{t}(y)a_{x}^{*}a_{y}^{*}+\overline{\varphi_{t}(x)}\overline{\varphi_{t}(y)}a_{x}a_{y}\right)\\
		& \quad+\frac{\lambda}{\sqrt N}\int_{\bR^3\times\bR^3}\dx\dy\,w(x-y)a_{x}^{*}\left(\overline{\varphi_{t}(y)}a_{y}\chi(\mathcal{N}\leq M)+\varphi_{t}(y)\chi(\mathcal{N}\leq M)a_{y}^{*}\right)a_{x}\\
		& \quad+\frac{\lambda}{2N}\int_{\bR^3\times\bR^3}\dx\dy\,w(x-y)a_{x}^{*}a_{y}^{*}a_{y}a_{x}.
	\end{align*}
	We remark that $M$ will be chosen to be $M=N$ later in the proof of Lemma \ref{lem:NjU}. Define a unitary operator $\cU^{(M)}_{N}$ by
	\begin{equation}\label{eq:def_mathcalUM}
		\mathrm{i}\partial_{t}\cU^{(M)}_{N}\left(t;s\right)=\mathcal{L}^{(M)}_N(t)\cU^{(M)}_{N}(t;s)\quad\text{and}\quad\cU^{(M)}_{N}\left(s;s\right)=1.
	\end{equation}
	
	\begin{lemma}\label{lem:UMNUM}
		Suppose that the assumptions in Theorem \ref{thm:main} hold and {let} $\cU^{(M)}_{N}$ be {the} unitary operator defined in \eqref{eq:def_mathcalUM}. Then, for any $j\in\mathbb N$ there exists a constant $K=K(C_0,\|V\|_{W^{2,\infty}}, C^V, \|w\|_1, \|w\|_2, j)>0$ such that for all $N\in\mathbb{N}$, $M>0$, $\psi\in\mathcal{F}$, and $t,s\in\mathbb{R}$,
		\[
		\left\langle \cU^{(M)}_{N}(t;s)\psi,(\cN+1)^{j}\cU^{(M)}_{N}(t;s)\psi\right\rangle\leq C\left\langle \psi,(\cN+1)^{j}\psi\right\rangle\exp\left(K\,(1+\sqrt{M/N})\right).
		\]
	\end{lemma}
	
	\begin{proof}
		Following the proof of \cite[Lemma 3.5]{RodnianskiSchlein2009}, see \cite[(3.15)]{RodnianskiSchlein2009}, we have
		\begin{equation}\label{eq:ddtUM}
			\frac{\mathrm{d}}{\dt}\langle \cU^{(M)}_{N}(t;0)\psi,(\cN+1)^{j}\cU^{(M)}_{N}(t;0)\psi\rangle=\lambda(A+B),
		\end{equation}
		where
		\begin{equation*}
			\begin{split}
				A
				&:=2\sum_{k=0}^{j-1}{\binom{j}{k}}(-1)^{k}\ImPart \int_{\bR^3\times\bR^3}\dx\dy\,w(x-y)\varphi_{t}(x)\varphi_{t}(y) \\
				&\qquad\times\langle \cU^{(M)}_{N}(t;0)\psi,\left(\cN^{k/2}a_{x}^{*}a_{y}^{*}(\cN+2)^{k/2}+(\cN+1)^{k/2}a_{x}^{*}a_{y}^{*}(\cN+3)^{k/2}\right)\cU^{(M)}_{N}(t;0)\psi\rangle\\
				B&:=\frac{2}{\sqrt{N}} \sum_{k=0}^{j-1}{\binom{j}{k}}\ImPart \int_{\bR^3}\dx\,
				\langle \cU^{(M)}_{N}(t;0)\psi,a_{x}^{*}a(w\left(x-\cdot)\varphi_t\right)\chi(\cN \leq M)(\cN+1)^{k/2}a_{x}\cN^{k/2}\cU^{(M)}_{N}(t;0)\psi\rangle.
			\end{split}
		\end{equation*}
		To control the contribution from the first term on the right-hand
		side of \eqref{eq:ddtUM}, we use bounds of the form
		\begin{align*}
			& \left|\int_{\bR^3\times\bR^3}\dx\dy\,w(x-y)\varphi_{t}(x)\varphi_{t}(y)\langle\cU^{(M)}_{N}(t;0)\psi,(\cN+1)^{\frac{k}{2}}a_{x}^{*}a_{y}^{*}(\cN+3)^{\frac{k}{2}}\cU^{(M)}_{N}(t;0)\psi\rangle \right|\\
			& \leq\int_{\bR^3\times\bR^3}\dx\dy\,\left|w(x-y)\varphi_{t}(x)\varphi_{t}(y)\right|\|a_{x}(\cN+1)^{\frac{k}{2}}\cU^{(M)}_{N}(t;0)\psi\|\,\|a_{y}^{*}(\cN+3)^{\frac{k}{2}}\cU^{(M)}_{N}(t;0)\psi\|\\
			& \leq\|w\|_{1}\|\varphi_{t}\|_{\infty}^{2}\left(\int_{\bR^3}\dx\,\|a_{x}(\cN+1)^{\frac{k}{2}}\cU^{(M)}_{N}(t;0)\psi\|^{2}\right)^{1/2}\left(\int_{\bR^3}\dy\,\|a_{y}(\cN+3)^{\frac{k}{2}}\cU^{(M)}_{N}(t;0)\psi\|^{2}\right)^{1/2}\\
			& \leq\|w\|_{1}\|\varphi_{t}\|_{\infty}^{2}\|(\cN+3)^{\frac{k+1}{2}}\cU^{(M)}_{N}(t;0)\psi\|^{2}.
		\end{align*}
		Here we used Young's inequality in the second step and the definition of the number operator in \eqref{eq:number operator}.
		
		On the other hand, to control the second
		integral on the right-hand side of \eqref{eq:ddtUM}, we use that
		\begin{align*}
			&\left|\int_{\bR^3}\dx\,
			\langle \cU^{(M)}_{N}(t;0)\psi,a_{x}^{*}a(w\left(x-\cdot)\varphi_t\right)\chi(\cN \leq M)(\cN+1)^{k/2}a_{x}\cN^{k/2}\cU^{(M)}_{N}(t;0)\psi\rangle\right|\\
			& \quad\leq\int_{\bR^3}\dx\,\Vert a_{x}(\cN+1)^{\frac{k}{2}}\cU^{(M)}_{N}(t;0)\psi\Vert \Vert a(w(x-\cdot)\varphi_{t})\chi(\cN \leq M)\Vert \Vert a_{x}\cN^{\frac{k}{2}}\cU^{(M)}_{N}(t;0)\psi\Vert \\
			& \quad\leq M^{1/2}\sup_{x}\|w(x-\cdot)\varphi_{t}\|\|(\cN+1)^{\frac{k+1}{2}}\cU^{(M)}_{N}(t;0)\psi\|^{2}\\
			& \quad\leq M^{1/2}\|w\|_{2}\,\|\varphi_{t}\|_{\infty}\,\|(\cN+1)^{\frac{k+1}{2}}\cU^{(M)}_{N}(t;0)\psi\|^{2}
		\end{align*}
		Using $ |\lambda |\le1$, this gives us that
		\begin{align*}
			&\left\vert\frac{\mathrm{d}}{\dt}\langle \cU^{(M)}_{N}(t;0)\psi,(\cN+1)^{j}\cU^{(M)}_{N}(t;0)\psi\rangle\right\vert\\
			&\le 2\sum_{k=0}^{j-1}{\binom{j}{k}}2\|w\|_{1}\|\varphi_{t}\|_{\infty}^{2}\|(\cN+3)^{\frac{k+1}{2}}\cU^{(M)}_{N}(t;0)\psi\|^{2}\\
			&\qquad+\frac{2}{\sqrt{N}} \sum_{k=0}^{j-1}{\binom{j}{k}}M^{1/2}\|w\|_{2}\,\|\varphi_{t}\|_{\infty}\,\|(\cN+1)^{\frac{k+1}{2}}\cU^{(M)}_{N}(t;0)\psi\|^{2}\\
			&\le 2 \sum_{k=0}^{j-1}{\binom{j}{k}}\left(2\|w\|_{1}\|\varphi_{t}\|_{\infty}^{2}+\frac{1}{\sqrt{N}}M^{1/2}\|w\|_{2}\,\|\varphi_{t}\|_{\infty}\right)\|\left(\cN+3\right)^{\frac{j}{2}}\cU^{(M)}_{N}(t;0)\psi\|^{2}\\
			&\le 2\cdot2^j\left(2\|w\|_{1}\|\varphi_{t}\|_{\infty}^{2}+\Big(\frac{M}{N}\Big)^{1/2}\|w\|_{2}\,\|\varphi_{t}\|_{\infty}\right)\Big\|\left(3\left(\cN+1\right)\right)^{\frac{j}{2}}\cU^{(M)}_{N}(t;0)\psi\Big\|^{2}\\
			&\le 2\cdot6^j\left(2\|w\|_{1}\|\varphi_{t}\|_{\infty}^{2}+\Big(\frac{M}{N}\Big)^{1/2}\|w\|_{2}\,\|\varphi_{t}\|_{\infty}\right)\Big\langle \cU^{(M)}_{N}(t;0)\psi,(\cN+1)^{j}\cU^{(M)}_{N}(t;0)\psi\Big\rangle.
		\end{align*}
		Applying the Grönwall lemma together with Proposition \ref{prop:dispersive d>=3 small}, we get 
		\begin{align*}
			&\Big\langle \cU^{(M)}_{N}(t;0)\psi,(\cN+1)^{j}\cU^{(M)}_{N}(t;0)\psi\Big\rangle\\
			&\leq \langle \psi,(\cN+1)^{j}\psi\rangle\exp\left(\int_0^t\dr s 2\cdot6^j\left(2\|w\|_{1}\|\varphi_{s}\|_{\infty}^{2}+\Big(\frac{M}{N}\Big)^{1/2}\|w\|_{2}\,\|\varphi_{s}\|_{\infty}\right)\right)\\
			&\leq \langle \psi,(\cN+1)^{j}\psi\rangle\exp\left(\int_0^t\dr s 2\cdot6^j\left(2\|w\|_{1}C_0^2(1+|s|)^{-3}+\Big(\frac{M}{N}\Big)^{1/2}\|w\|_{2}\,C_0(1+|s|)^{-\frac{3}{2}}\right)\right)\\
			&\leq \langle \psi,(\cN+1)^{j}\psi\rangle\exp\left(K\left(1+\Big(\frac{M}{N}\Big)^{1/2}\right)\right).
		\end{align*}
	This gives us the desired lemma.
	\end{proof}
	
	\section{Comparison of Dynamics}\label{sec:comparison-of-dynamics}
	
	The main goal of this section is to provide important lemmata to prove Proposition \ref{prop:Et1} and \ref{prop:Et2}.
	
	\begin{lemma}\label{lem:NjU}
		Suppose that the assumptions in Theorem \ref{thm:main} hold. Let $\cU\left(t;s\right)$ be the unitary evolution defined in \eqref{eq:def_mathcalU}. Then for any $\psi\in\mathcal{F}$ and $j\in\mathbb{N}$, there exists a constant $C \equiv C(j, C_0, \|V\|_{W^{2,\infty}}, C^V, \|w\|_1, \|w\|_2)>0$ such that
		\[
		\langle \cU(t;s)\psi,\cN^{j}\cU(t;s)\psi\rangle\leq C\langle \psi,(\cN+1)^{2j+2}\psi\rangle,
		\]
	\end{lemma}
	
	In this section we modify the proof given in previous articles, for example, \cite{RodnianskiSchlein2009} or \cite{ChenLeeLee2018} and the references therein.
	In \cite{ChenLeeLee2018,RodnianskiSchlein2009}, the Hardy inequality was used to cover singular interaction potentials. However, here we use H\"older's inequality and Proposition \ref{prop:dispersive d>=3 small}. 

	We now begin the proof of Lemma \ref{lem:NjU}. 
	To prove the lemma, we compare the dynamics of $\cU$ and $\cU^{(M)}$ in Lemma \ref{lem:UNUUM}.
	To do so, we recall weak bounds on the $\cU$ dynamics. 
	
	\begin{lemma}[{\cite[Lemma 3.6]{RodnianskiSchlein2009}} ] 
		\label{lem:3.6 i Rodnianski}
		For arbitrary $t,s\in\mathbb{R}$ and $\psi\in\mathcal{F}$, we have {
			\[
			\left\langle \psi,\cU(t;s) \mathcal{N}\cU^{*}(t;s)\psi\right\rangle \leq6\left\langle \psi,(\mathcal{N}+N+1)\psi\right\rangle .
			\] }
		Moreover, for every $\ell\in\mathbb{N}$, there exists a constant $C(\ell)$ such that
		\[
		\left\langle \psi,\cU(t;s)\mathcal{N}^{2\ell}\cU^{*}(t;s)\psi\right\rangle \leq C(\ell)\left\langle \psi,(\mathcal{N}+N)^{2\ell}\psi\right\rangle ,
		\]
		\[
		\left\langle \psi,\cU(t;s)\mathcal{N}^{2\ell+1}\cU^{*}(t;s)\psi\right\rangle \leq C(\ell)\left\langle \psi,(\mathcal{N}+N)^{2\ell+1}(\mathcal{N}+1)\psi\right\rangle
		\]
		for all $t,s\in\mathbb{R}$ and $\psi\in\mathcal{F}$.
	\end{lemma}
	
	\begin{proof}
		The proof can be found in \cite{RodnianskiSchlein2009}.
	\end{proof}

	Now we are ready to compare the dynamics of $\cU$ and $\cU^{(M)}_{N}$.
	
	\begin{lemma}\label{lem:UNUUM}
		Suppose that the assumptions in Theorem \ref{thm:main} hold. Then, for every $j\in\mathbb{N}$ and $\psi\in\cF$, there exists a constant $C \equiv C(j, C_0, \|V\|_{W^{2,\infty}}, C^V, \|w\|_1, \|w\|_2 )>0$ such that 
		for all $t,s\in\mathbb R$
		\begin{align}
			& \Big|\langle \cU(t;s)\psi,\cN^{j}\big(\cU(t;s)-\cU^{(M)}_{N}(t;s)\big)\psi\rangle\Big|\nonumber\\
			& \quad\leq C(N/M)^{j}\|(\cN+1)^{j+1}\psi\|^{2}\exp\Big(K\big(1+\sqrt{M/N}\big)\Big)
		\end{align}
		and
		\begin{equation}
			\begin{split}
				&\Big| \langle \cU^{(M)}_N (t;0) \psi, \cN^j \left( \cU (t;0) -
				\cU_N^{(M)} (t;0)\right) \psi \rangle \Big|
				\\
				&\leq \; C(N/M)^{j} \|  (\cN+1)^{j+1}
				\psi \|^2\exp\Big(K\big(1+\sqrt{M/N}\big)\Big).
			\end{split}\label{eq:UNUUM1overM}
		\end{equation}
	\end{lemma}
	
	\begin{proof}
		To simplify the notation, we consider the case $s=0$ and $t>0$ only;
		other cases can be treated in a similar manner. To prove the first
		inequality of the lemma, we expand the difference of the two evolutions
		as follows:
		\begin{align*}
			& \langle \cU(t;0)\psi,\cN^{j}\big(\cU(t;0)-\cU^{(M)}_{N}(t;0)\big)\psi\rangle=\langle \cU(t;0)\psi,\cN^{j}\cU(t;0)\big(1-\cU^{*}(t;0)\cU^{(M)}_{N}(t;0)\big)\psi\rangle\\
			& =-\ii \int_{0}^{t}\ds\,\langle \cU(t;0)\psi,\cN^{j}\cU(t;0)\big(\cL_{N}(s)-\cL_{N}^{(M)}(s)\big)\cU^{(M)}_{N}(s;0)\psi\rangle\\
			& =-\frac{\ii \lambda }{\sqrt{N}}\int_{0}^{t}\ds\,\int_{\bR^3\times\bR^3}\dx\dy\,w(x-y)\\
			& \qquad\times\langle\cU(t;0)\psi,\cN^{j}\cU(t;s)a_{x}^{*}\big(\overline{\varphi_{s}(y)}a_{y}\chi(\cN >M)+\varphi_{s}(y)\chi(\cN >M)a_{y}^{*}\big)a_{x}\cU^{(M)}_{N}(s;0)\psi\rangle \\
			& =-\frac{\ii \lambda }{\sqrt{N}}\int_{0}^{t}\ds\,\int_{\bR^3}\dx\,\langle a_{x}\cU^{*}(t;s)\cN^{j}\cU(t;0)\psi,a(w(x-\cdot)\varphi_{s})\chi(\cN >M)a_{x}\cU^{(M)}_{N}(s;0)\psi\rangle \\
			& \quad\quad-\frac{\ii \lambda }{\sqrt{N}}\int_{0}^{t}\ds\,\int_{\bR^3}\dx\,\langle a_{x}\cU^{*}(t;s)\cN^{j}\cU(t;0)\psi,\chi(\cN >M)a^{*}(w(x-\cdot)\varphi_{s})a_{x}\cU^{(M)}_{N}(s;0)\psi\rangle .
		\end{align*}
		Hence, using $|\lambda|\le1$,
		\begin{align}
			& \Big|\langle \cU(t;0)\psi,\cN^{j}\big(\cU(t;0)-\cU^{(M)}_{N}(t;0)\big)\psi\rangle\Big|\nonumber \\
			& \leq\frac{1}{\sqrt{N}}\int_{0}^{t}\ds\,\int_{\bR^3}\dx\,\|a_{x}\cU^{*}(t;s)\cN^{j}\cU(t;0)\psi\|\nonumber\\
			&\qquad\qquad \times\|a(w(x-\cdot)\varphi_{s})a_{x}\chi(\cN >M+1)\cU^{(M)}_{N}(s;0)\psi\|\nonumber \\
			& \quad+\frac{1}{\sqrt{N}}\int_{0}^{t}\ds\,\int_{\bR^3}\dx\,\|a_{x}\cU^{*}(t;s)\cN^{j}\cU(t;0)\psi\|\nonumber\\
			&\qquad\qquad \times\|a^{*}(w(x-\cdot)\varphi_{s})a_{x}\chi(\cN >M)\cU^{(M)}_{N}(s;0)\psi\|\nonumber \\
			& \leq\frac{2}{\sqrt{N}}\int_{0}^{t}\ds\,\sup_{x}\|w(x-\cdot)\varphi_{s}\|\nonumber\\
			&\qquad\qquad \times \int_{\bR^3}\dx\,\|a_{x}\cU^{*}(t;s)\cN^{j}\cU(t;0)\psi\|\,\|a_{x}(\cN+1)^{1/2}\chi(\cN >M)\cU^{(M)}_{N}(s;0)\psi\|\nonumber \\
			& \leq\frac{2}{\sqrt{N}}\int_{0}^{t}\ds\,\sup_{x}\|w(x-\cdot)\varphi_{s}\|\nonumber\\
			&\qquad\qquad \times \|\cN^{1/2}\cU^{*}(t;s)\cN^{j}\cU(t;0)\psi\|\,\|(\cN+1)\chi(\cN >M)\cU^{(M)}_{N}(s;0)\psi\|.\label{eq:U-UM}
		\end{align}
		Since $\chi(\cN >M)\leq(\cN /M)^{L}$ for any $L>1$,
		we find that, from \cite[(3.30)]{RodnianskiSchlein2009},
		\begin{align*}
			& \Big|\langle \cU(t;0)\psi,\cN^{j}\big(\cU(t;0)-\cU^{(M)}_{N}(t;0)\big)\psi\rangle\Big|\\
			& \quad\leq C N^{j}\|(\cN+1)^{j+1}\psi\|\nonumber\\
			&\qquad\qquad \times \int_{0}^{t}\ds\,\sup_{x}\|w(x-\cdot)\varphi_{s}\|_{2}\langle \cU^{(M)}_{N}(s;0)\psi,(\cN+1)^{2}\chi(\cN >M)\cU^{(M)}_{N}(s;0)\psi\rangle ^{1/2}\\
			& \quad\leq C N^{j}\|(\cN+1)^{j+1}\psi\|\nonumber\\
			&\qquad\qquad \times \int_{0}^{t}\ds\,\|w\|_{2}\|\varphi_{s}\|_{\infty}\langle \cU^{(M)}_{N}(s;0)\psi,(\cN+1)^{2}\chi(\cN >M)\cU^{(M)}_{N}(s;0)\psi\rangle ^{1/2}\\
			& \quad\leq C N^{j}\|(\cN+1)^{j+1}\psi\|\langle \cU^{(M)}_{N}(s;0)\psi,\frac{(\cN+1)^{2+2j}}{M^{2j}}\cU^{(M)}_{N}(s;0)\psi\rangle ^{1/2}{\int_{0}^{t}\ds\,\|\varphi_{s}\|_{\infty}}
		\end{align*}
		where
		\[
		C = C(j,\|V\|_{W^{2,\infty}}, C^V, \|w\|_1, \|w\|_2 ).
		\]
		{By Lemma \ref{lem:UMNUM} and $\|\varphi_s\|_\infty \leq C(1+s)^{3/2}$, see Proposition \ref{prop:dispersive d>=3 small}}, we conclude that
		\begin{align*}
			& \Big|\langle \cU(t;0)\psi,\cN^{j}\big(\cU(t;0)-\cU^{(M)}_{N}(t;0)\big)\psi\rangle\Big|\\
			& {\quad\leq C(N/M)^{j}\|(\cN+1)^{j+1}\psi\|^{2}\exp\Big(K \big(1+\sqrt{M/N}\big)\Big)\int_{0}^{t}\ds\,(1+s)^{3/2}}\\
			& \quad\leq C(N/M)^{j}\|(\cN+1)^{j+1}\psi\|^{2}\exp\Big(K \big(1+\sqrt{M/N}\big)\Big)\\
			& \quad\leq C(N/M)^{j}\|(\cN+1)^{j+1}\psi\|^{2}\exp\Big(K\big(1+\sqrt{M/N}\big)\Big).\\
		\end{align*}
		To prove \eqref{eq:UNUUM1overM}, we proceed similarly; analogously to \eqref{eq:U-UM} we find
		\begin{equation*}
			\begin{split}
				&\langle \cU^{(M)}_N (t;0) \psi, \cN^j \left( \cU (t;0) -
				\cU_N^{(M)}
				(t;0)\right) \psi \rangle \\
				= \; &-\frac{\ii\lambda }{\sqrt{N}}\int_0^t \dr s \int_{\bR^3} \dr x  \langle a_x
				\cU (t;s)^* \cN^j \cU^{(M)}_N (t;0) \psi, a (w(x-.)\varphi_t) \chi
				(\cN > M) a_x \cU_N^{(M)} (s;0) \psi \rangle \\
				&-\frac{\ii\lambda }{\sqrt{N}}\int_0^t \dr s \int_{\bR^3} \dr x  \langle a_x \cU
				(t;s)^* \cN^j \cU^{(M)}_N (t;0) \psi, \chi (\cN >M) a^*
				(w(x-.)\varphi_t) a_x \cU_N^{(M)} (s;0) \psi \rangle
			\end{split}
		\end{equation*}
		and thus, by $|\lambda|\le1$, $\chi (\cN > M)\le\left(\frac{\cN}{M}\right)^{L}$ for any $L>1$ and Proposition \ref{prop:dispersive d>=3 small},
		\begin{equation}\label{eq:0}
			\begin{split}
				&\Big| \langle \cU^{(M)}_N (t;0) \psi, \cN^j \left( \cU (t;0) -
				\cU_N^{(M)} (t;0)\right) \psi \rangle \Big| \\
				&\leq \; \frac{C}{\sqrt{N}} \int_0^t \dr s\,\sup_{x}\|w(x-\cdot)\varphi_{s}\|_{2}\, \|\cN^{1/2} \cU
				(t;s)^* \cN^j \cU_N^{(M)} (t;0) \psi \| \, \|  \cN \chi (\cN > M)
				\cU_N^{(M)} (s;0) \psi \|\,
				\\
				&\leq \; \frac{C}{\sqrt{N}} \int_0^t \dr s\,\|w\|_{2}\, \|\varphi_{s}\|_{\infty}\, \|\cN^{1/2} \cU
				(t;s)^* \cN^j \cU_N^{(M)} (t;0) \psi \| \, \|  \cN \left(\frac{\cN}{M}\right)^{j}
				\cU_N^{(M)} (s;0) \psi \|\,
				\\
				&\leq \; \frac{C}{M^{j}\sqrt{N}}\|w\|_{2}\, \int_0^t \dr s\, C_0(1+|s|)^{-3/2}\, \|\cN^{1/2} \cU
				(t;s)^* \cN^j \cU_N^{(M)} (t;0) \psi \| \, \|  \cN^{j+1}
				\cU_N^{(M)} (s;0) \psi \|\,.
			\end{split}
		\end{equation}
		By Lemma \ref{lem:3.6 i Rodnianski}, Lemma \ref{lem:UMNUM} and $j+1/2\le j+1$ , we have
		\begin{equation}\label{eq:1}
			\begin{split}
				& \|\cN^{1/2} \cU
				(t;s)^* \cN^j \cU_N^{(M)} (t;0) \psi \| \le 6 \|(\cN+N+1)^{1/2}  \cN^j \cU_N^{(M)} (t;0) \psi \| \\
				& \le 12 N^{j+1/2} \|(\cN+1)^{j+1/2}  \cU_N^{(M)} (t;0) \psi \| \\
				& \le C N^{j+1/2} \|(\cN+1)^{j+1/2}   \psi \| \exp\Big(K\big(1+\sqrt{M/N}\big)\Big) \\
				& \le C N^{j+1/2} \|(\cN+1)^{j+1}   \psi \| \exp\Big(K\big(1+\sqrt{M/N}\big)\Big).
			\end{split}
		\end{equation}
		By Lemma \ref{lem:UMNUM}, we have
		\begin{equation}\label{eq:2}
			\|  \cN^{j+1}
			\cU_N^{(M)} (s;0) \psi \|\le C\|  (\cN+1)^{j+1}
			\psi \|\exp\Big(K\big(1+\sqrt{M/N}\big)\Big).
		\end{equation}
		Combining \eqref{eq:0}, \eqref{eq:1} and \eqref{eq:2}, we obtain
		\begin{equation}
			\begin{split}
				&\Big| \langle \cU^{(M)}_N (t;0) \psi, \cN^j \left( \cU (t;0) -
				\cU_N^{(M)} (t;0)\right) \psi \rangle \Big|
				\\
				&\leq \; \frac{C}{M^{j}\sqrt{N}}\, N^{j+1/2} \|  (\cN+1)^{j+1}
				\psi \|^2\exp\Big(2K\big(1+\sqrt{M/N}\big)\Big)\\
				&\leq \; \frac{C}{M^{j}}\, N^{j} \|  (\cN+1)^{j+1}
				\psi \|^2\exp\Big(K\big(1+\sqrt{M/N}\big)\Big)\\
				&\leq \; C(N/M)^{j} \|  (\cN+1)^{j+1}
				\psi \|^2\exp\Big(K\big(1+\sqrt{M/N}\big)\Big).
			\end{split}
		\end{equation}Again, we have
		\[
		C = C(j, C_0, \|V\|_{W^{2,\infty}}, C^V, \|w\|_1, \|w\|_2 ).
		\]
		This gives us the desired lemma.
	\end{proof}
	
	Let us now prove Lemma \ref{lem:NjU}.
	\begin{proof}[Proof of Lemma \ref{lem:NjU}]
		Let $M=N$. Then by Lemmata \ref{lem:UMNUM} and \ref{lem:UNUUM}, we get
		\begin{align*}
			& \Big\langle \cU\left(t;s\right)\psi,\cN^{j}\cU\left(t;s\right)\psi\Big\rangle\\
			& =\Big\langle \cU\left(t;s\right)\psi,\cN^{j}(\cU-\cU^{(M)}_{N})\left(t;s\right)\psi\Big\rangle +\Big\langle (\cU-\cU^{(M)}_{N})\left(t;s\right)\psi,\cN^{j}\cU^{(M)}_{N}\left(t;s\right)\psi\Big\rangle\\
			& \qquad+\Big\langle \cU^{(M)}_{N}\left(t;s\right)\psi,\cN^{j}\cU^{(M)}_{N}\left(t;s\right)\psi\Big\rangle\\
			& \leq C\Big\langle \psi,\left(\cN+1\right)^{2j+2}\psi\Big\rangle.
		\end{align*}
		This leads us the desired result.
	\end{proof}
	
	Recall the definition of $\widehat{\cU}\left(t;s\right)$ in \eqref{eq:def_mathcalwidehatU}. In the next lemma, we prove an estimate for the evolution with respect to $\widehat{\cU}$.
	
	\begin{lemma}\label{lem:tildeNj}
		Suppose that the assumptions in Theorem \ref{thm:main} hold. Then, for any $\psi\in\cF$ and $j\in \bN$, there exists a constant $C = C(C_0, \|V\|_{W^{2,\infty}}, C^V, \|w\|_1 )>0$ such that
		\[
		\langle \widehat{\cU}\left(t;s\right)\psi,\cN^{j}\widehat{\cU}\left(t;s\right)\psi\rangle\leq C\langle \psi,\left(\cN+1\right)^{j}\psi\rangle . 
		\]
	\end{lemma}
	
	\begin{proof}
		Let $\widehat{\psi}=\widehat{\cU}(t;s)\psi$ and assume without loss of generality that $t\ge s$. We have
		\begin{align*}
			& \frac{\mathrm{d}}{\dt}\langle \widehat{\psi},(\cN+1)^{j}\widehat{\psi}\rangle
			=\langle \widehat{\psi},[\ii (\cL_{2}+\cL_{4}),(\cN+1)^{j}]\widehat{\psi}\rangle\\
			& ={-}{  \frac{1}{ N} }\ImPart \int_{\bR^3\times\bR^3}\dx\dy\,w(x-y)\varphi_{t}(x)\varphi_{t}(y)\langle \widehat{\psi},[a_{x}^{*}a_{y}^{*},(\cN+1)^{j}]\widehat{\psi}\rangle\\
			& ={  \frac{1}{ N} }\ImPart \int_{\bR^3\times\bR^3}\dx\dy\,w(x-y)\varphi_{t}(x)\varphi_{t}(y)\langle \widehat{\psi},a_{x}^{*}a_{y}^{*}((\cN+3)^{j}-(\cN+1)^{j})\widehat{\psi}\rangle\\
			& ={  \frac{1}{ N} }\ImPart \int_{\bR^3\times\bR^3}\dx\dy\,w(x-y)\varphi_{t}(x)\varphi_{t}(y)\\
			& \qquad\times\langle (\cN+3)^{(j+1)/2}a_{x}\widehat{\psi},a_{y}(\cN+3)^{(1-j)/2}((\cN+3)^{j}-(\cN+1)^{j})\widehat{\psi}\rangle.
		\end{align*}
		Then, using Young’s inequality, one gets
		\begin{align*}
			\frac{\mathrm{d}}{\dt}\langle \widehat{\psi},(\cN+1)^{j}\widehat{\psi}\rangle
			&\leq {  \frac{1}{ N} }\int_{\bR^3\times\bR^3}\dx\dy\,|w(x-y)| |\varphi_{t}(x)| |\varphi_{t}(y)|\|(\cN+3)^{(j-1)/2}a_{x}\widehat{\psi}\|\\
			& \qquad\times\|a_{y}(\cN+3)^{(1-j)/2}((\cN+3)^{j}-(\cN+1)^{j})\widehat{\psi}\|\\
			& \leq {  \frac{1}{ N} }\|\varphi_{t}\|_{\infty}^{2} \Big(\int_{\bR^3}\dx\,\|(\cN+3)^{(j-1)/2}a_{x}\widehat{\psi}\|^{2}\Big)^{1/2}\\
			& \qquad\times\Big(\int_{\bR^3}\dy\,\|a_{y}(\cN+3)^{(1-j)/2}((\cN+3)^{j}-(\cN+1)^{j})\widehat{\psi}\|^2\Big)^{1/2}
		\end{align*}
		Since, for all $j\in\bN$, $|(\cN+3)^{j}-(\cN+1)^{j})|\leq C_j (\cN+1)^{j-1}$ for some $C_j>0$ ,
		we have that
		\begin{align*}
			\frac{\mathrm{d}}{\dt}\langle \widehat{\cU}(t;s)\psi,(\cN+1)^{j}\widehat{\cU}(t;s)\psi\rangle
			&\leq {  \frac{C_j}{ N} }\|\varphi_{t}\|_{\infty}^{2}\|(\cN+1)^{j/2}\widehat{\cU}(t;s)\psi\|^{2}\\
			&={  \frac{C_j}{ N} }\|\varphi_{t}\|_{\infty}^{2}\langle \widehat{\cU}(t;s)\psi,(\cN+1)^{j}\widehat{\cU}(t;s)\psi\rangle.
		\end{align*}
		Here, note that $C_j$ can change from line to line.
		Applying Grönwall's lemma with Proposition \ref{prop:dispersive d>=3 small}, we conclude that
		\begin{align*}
			\langle \widehat{\cU}(t;s)\psi,(\cN+1)^{j}\widehat{\cU}(t;s)\psi\rangle
			&\leq \langle \psi,(\cN+1)^{j}\psi\rangle \,
			\exp\Big(\int_s^t \mathrm{d} r\, \frac{C_j}{N}C_0^2 \frac{1}{(1+|r|)^3}\Big)\\
			&\leq C\langle \psi,(\cN+1)^{j}\psi\rangle.
		\end{align*}
		Hence, we get the result.
	\end{proof}
	
	\begin{lemma}
		For all $\psi\in\cF$ and $f\in L^2(\bR^3)$,
		we have the following inequalities with a constant $C = C(C_0,  \|V\|_{W^{2,\infty}}, C^V, \|w\|_1, \|w\|_2 ) > 0$.
		Then we have
		\begin{align}
			&\Vert (\cN+1)^{j/2}\cL_{3}(t)\psi\Vert \leq
			\frac{C}{N} \|\varphi_{t}\|_{L^{\infty}}\Vert \left(\cN+1\right)^{(j+3)/2}\psi\Vert,\label{eq:NL3-0}\\
			&\Vert (\cN+1)^{j/2}\big(\cU^{*}(t;0)\phi(f)\cU(t;0)-\widehat{\cU}^{*}(t;0)\phi(f)\widehat{\cU}(t;0)\big)\Omega\Vert \leq\frac{C\|f\|}{N}.\label{eq:NjUphiUtildeUphitildeU-0}
		\end{align}
	\end{lemma}
	
	\begin{proof}
		For \eqref{eq:NL3-0}, we copy the proof from \cite[Lemma 5.3]{Lee2013} with $\|w\|_2 < \infty$.
		
		For \eqref{eq:NjUphiUtildeUphitildeU-0}, we follow the proof from \cite[Lemma 5.4]{Lee2013} 
		with Lemma \ref{lem:tildeNj} and \eqref{eq:NL3-0}.
	\end{proof}
	
	\begin{proof}[Proof of Proposition \ref{prop:Et1}]
		Recall that
		\[
		E_{t}^{(1)}(J)=\frac{d_{N}}{N}\Big\langle W^{*}(\sqrt{N}\varphi_0)\frac{(a^{*}(\varphi_0))^{N}}{\sqrt{N!}}\Omega,\cU^{*}(t;0)d\Gamma(J)\cU(t;0)\Omega\Big\rangle.
		\]
		We begin by noting that
		\begin{align*}
			|E_{t}^{(1)}(J)| & =\Big|\frac{d_{N}}{N}\langle W^{*}(\sqrt{N}\varphi_0)\frac{(a^{*}(\varphi_0))^{N}}{\sqrt{N!}}\Omega,\cU^{*}(t;0)d\Gamma(J)\cU(t;0)\Omega\rangle\Big|\\
			& \leq\frac{d_{N}}{N}\Big\Vert (\cN+1)^{-\frac{1}{2}}W^{*}(\sqrt{N}\varphi_0)\frac{(a^{*}(\varphi_0))^{N}}{\sqrt{N!}}\Omega\Big\Vert\,\Big\Vert (\cN+1)^{\frac{1}{2}}\cU^{*}(t;0)d\Gamma(J)\cU(t;0)\Omega\Big\Vert .
		\end{align*}
		Recall that, by Lemma \ref{lem:coherent_all}, 
		\[
		\Vert (\cN+1)^{-\frac{1}{2}}W^{*}(\sqrt{N}\varphi_0)\frac{(a^{*}(\varphi_0))^{N}}{\sqrt{N!}}\Omega\Vert \leq\frac{C}{d_{N}}
		\]
		and, by applying Lemma \ref{lem:NjU} twice,
		\begin{align*}
			& \Vert (\cN+1)^{\frac{1}{2}}\cU^{*}(t;0)d\Gamma(J)\cU(t;0)\Omega\Vert 
			\leq C\Vert (\cN+1)^{j+1}d\Gamma(J)\cU(t;0)\Omega\Vert \\
			& \qquad\leq C\Vert J\Vert \, \Vert (\cN+1)^{j+2}\cU(t;0)\Omega\Vert 
			\leq C\Vert J\Vert \, \Vert (\cN+1)^{j+3}\Omega\Vert=C\|J\|,
		\end{align*}
		we obtain
		\[
		|E_{t}^{(1)}(J)|\leq\frac{C\|J\|}{N},
		\]
		which is the desired result.
	\end{proof}

	\begin{proof}[Proof of Proposition \ref{prop:Et2}]
		Let
		\[
		\cR (f)=\cU^{*}(t;0)\phi(f)\cU(t;0)-\widehat{\cU}^{*}(t;0)\phi(f)\widehat{\cU}(t;0).
		\]
		Then
		\begin{align}
			&|E_{t}^{(2)}(J)|\nonumber\\
			& =\frac{d_{N}}{\sqrt{N}} \Big\langle W^{*}(\sqrt{N}\varphi_0)\frac{(a^{*}(\varphi_0))^{N}}{\sqrt{N!}}\Omega,\mathcal{\widehat{U}}^{*}(t;0)\phi(J\varphi_{t})\mathcal{\widehat{U}}(t;0)\Omega\Big\rangle\nonumber \\
			& \qquad+\frac{d_{N}}{\sqrt{N}}\Big\langle W^{*}(\sqrt{N}\varphi_0) \frac{(a^{*}(\varphi_0))^{N}}{\sqrt{N!}}\Omega,\cR (J\varphi_{t})\Omega\Big\rangle\nonumber \\
			& \leq \frac{d_{N}}{\sqrt{N}} \Big\Vert \sum_{k=0}^{\infty}(\cN+1)^{-\frac{5}{2}}P_{2k+1}W^{*}(\sqrt{N}\varphi_0)\frac{(a^{*}(\varphi_0))^{N}}{\sqrt{N!}}\Omega\Big\Vert \, \Big\Vert (\cN+1)^{\frac{5}{2}}\mathcal{\widehat{U}}^{*}(t;0)\phi(J\varphi_{t})\widehat{\cU}(t;0)\Omega\Big\Vert \nonumber \\
			& \qquad+\frac{d_{N}}{\sqrt{N}}\Big\Vert (\cN+1)^{-\frac{1}{2}}W^{*}(\sqrt{N}\varphi_0)\frac{(a^{*}(\varphi_0))^{N}}{\sqrt{N!}}\Omega\Big\Vert \, \Big\Vert (\cN+1)^{\frac{1}{2}}\cR (J\varphi_{t})\Omega \Big\Vert \label{eq:e_t^21}.
		\end{align}
		Let $\mathsf{K}=\frac{1}{2}N^{1/3}$. By Lemmata \ref{lem:coherent_all} and \ref{lem:coherent_even_odd}, we have
		\begin{align*}
			& \Big\Vert \sum_{k=0}^{\infty}(\cN+1)^{-\frac{5}{2}}P_{2k+1}W^{*}(\sqrt{N}\varphi_0)\frac{(a^{*}(\varphi_0))^{N}}{\sqrt{N!}}\Omega\Big\Vert^{2}\\
			& \qquad\leq\sum_{k=0}^{\mathsf{K}}\Big\Vert (\cN+1)^{-\frac{5}{2}}P_{2k+1}W^{*}(\sqrt{N}\varphi_0)\frac{(a^{*}(\varphi_0))^{N}}{\sqrt{N!}}\Omega\Big\Vert^{2}\\
			& \qquad\qquad+\frac{1}{\mathsf{K}^{4}}\sum_{k=\mathsf{K}}^{\infty}\Big\Vert(\cN+1)^{-1/2} P_{2k+1}W^{*}(\sqrt{N}\varphi_0)\frac{(a^{*}(\varphi_0))^{N}}{\sqrt{N!}}\Omega\Big\Vert^{2}\\
			& \qquad\leq \Big(\sum_{k=0}^{\mathsf{K}}\frac{C}{(k+1)^{2}d_{N}^{2}N}\Big)+\frac{C}{N^{4/3}}\Big\Vert(\cN+1)^{-1/2} W^{*}(\sqrt{N}\varphi_0)\frac{(a^{*}(\varphi_0))^{N}}{\sqrt{N!}}\Omega\Big\Vert^2 \leq\frac{C}{d_{N}^{2}N^{4/3}}.
		\end{align*}
		Using Lemma \ref{lem:tildeNj},
		\begin{alignat*}{1}
			& \Vert (\cN+1)^{\frac{5}{2}}\mathcal{\widehat{U}}^{*}(t;0)\phi(J\varphi_{t})\widehat{\cU}(t;0)\Omega\Vert 
			\leq C\Vert (\cN+1)^{\frac{5}{2}}\phi(J\varphi_{t})\widehat{\cU}(t;0)\Omega\Vert \\
			&\leq  C\Vert (\cN+1)^{3}\phi(J\varphi_{t})\widehat{\cU}(t;0)\Omega\Vert
			=  C\Vert \phi(J\varphi_{t}) (\cN+2)^{3}\widehat{\cU}(t;0)\Omega\Vert \\
			&\leq C\|J\varphi_{t}\|\Vert (\cN+2)^{7/2}\mathcal{\widehat{U}}(t;0)\Omega\Vert \leq C\|J\|\Vert (\cN+2)^{7/2}\Omega\Vert = C\|J\|.
		\end{alignat*}
		For the second term on the right-hand side of \eqref{eq:e_t^21}, we use Lemma \ref{lem:coherent_all}, and \eqref{eq:NjUphiUtildeUphitildeU-0}.
		for $f=J\varphi_t$.
		Altogether, we have
		\begin{equation*}
			\Vert (\cN+1)^{j/2}\cR (f)\Omega\Vert \leq C\|J\|N^{-1} 
		\end{equation*}
		which is the desired conclusion.
	\end{proof}


\begin{thebibliography}{10}
	
	\bibitem{wiemancornell}
	M.~H. Anderson, J.~R. Ensher, M.~R. Matthews, C.~E. Wieman, and E.~A. Cornell.
	\newblock Observation of {B}ose-{E}instein condensation in a dilute atomic
	vapor.
	\newblock {\em Science}, 269(5221):198--201, 1995.
	
	\bibitem{BardosGolseMauser2000}
	C.~Bardos, F.~Golse, and N.~J. Mauser.
	\newblock Weak coupling limit of the $ n $-particle {S}chr{\"o}dinger equation.
	\newblock {\em Methods and Applications of Analysis}, 7(2):275--294, 2000.
	
	\bibitem{NielsOliveiraSchlein2015}
	N.~Benedikter, G.~de~Oliveira, and B.~Schlein.
	\newblock Quantitative derivation of the {G}ross-{P}itaevskii equation.
	\newblock {\em Comm. Pure Appl. Math.}, 68(8):1399--1482, 2015.
	
	\bibitem{benedikter2016effective}
	N.~Benedikter, M.~Porta, and B.~Schlein.
	\newblock {\em Effective evolution equations from quantum dynamics}.
	\newblock Springer, 2016.
	
	\bibitem{boccato2017quantum}
	C.~Boccato, S.~Cenatiempo, and B.~Schlein.
	\newblock Quantum many-body fluctuations around nonlinear schr{\"o}dinger
	dynamics.
	\newblock In {\em Annales Henri Poincar{\'e}}, volume~18, pages 113--191.
	Springer, 2017.
	
	\bibitem{bose}
	{B}ose.
	\newblock Plancks gesetz und lichtquantenhypothese.
	\newblock {\em Zeit{S}chrift für Physik}, 26(1):178--181, 1924.
	
	\bibitem{brennecke2019fluctuations}
	C.~Brennecke, P.~T. Nam, M.~Napi{\'o}rkowski, and B.~Schlein.
	\newblock Fluctuations of n-particle quantum dynamics around the nonlinear
	schr{\"o}dinger equation.
	\newblock In {\em Annales de l'Institut Henri Poincar{\'e} C, Analyse non
		lin{\'e}aire}, volume~36, pages 1201--1235. Elsevier, 2019.
	
	\bibitem{BrenneckeSchlein2019GP}
	C.~Brennecke and B.~Schlein.
	\newblock Gross--pitaevskii dynamics for {B}ose--{E}instein condensates.
	\newblock {\em Analysis \& PDE}, 12(6):1513--1596, 2019.
	
	\bibitem{brscsc20222}
	C.~Brennecke, B.~Schlein, and S.~{S}chraven.
	\newblock Bogoliubov theory for trapped bosons in the {Gross}-{Pitaevskii}
	regime.
	\newblock {\em Ann. Henri Poincar{\'e}}, 23(5):1583--1658, 2022.
	
	\bibitem{brscsc2022}
	C.~Brennecke, B.~Schlein, and S.~{S}chraven.
	\newblock {B}ose-{{E}instein} condensation with optimal rate for trapped bosons
	in the {Gross}-{Pitaevskii} regime.
	\newblock {\em Math. Phys. Anal. Geom.}, 25(2):71, 2022.
	\newblock Id/No 12.
	
	\bibitem{cazenave}
	T.~Cazenave.
	\newblock {\em Semilinear {S}chrödinger Equations}.
	\newblock American Mathematical Society, 2003.
	
	\bibitem{cazenaveweissler}
	T.~Cazenave and F.~Weissler.
	\newblock Rapidly decaying solutions of the nonlinear {S}chrödinger equation.
	\newblock {\em Communications in Mathematical Physics}, 147:75–100, 1992.
	
	\bibitem{ChenLee2011}
	L.~Chen and J.~O. Lee.
	\newblock Rate of convergence in nonlinear {H}artree dynamics with factorized
	initial data.
	\newblock {\em J. Math. Phys.}, 52(5):052108, 25, 2011.
	
	\bibitem{ChenLeeLee2018}
	L.~Chen, J.~O. Lee, and J.~Lee.
	\newblock Rate of convergence toward {H}artree dynamics with singular
	interaction potential.
	\newblock {\em J. Math. Phys.}, 59(3):031902, 2018.
	
	\bibitem{ChenLeeSchlein2011}
	L.~Chen, J.~O. Lee, and B.~Schlein.
	\newblock Rate of convergence towards {H}artree dynamics.
	\newblock {\em J. Stat. Phys.}, 144(4):872--903, 2011.
	
	\bibitem{chen2011quintic}
	T.~Chen and N.~Pavlovi{\'c}.
	\newblock The quintic {NLS} as the mean field limit of a boson gas with
	three-body interactions.
	\newblock {\em Journal of Functional Analysis}, 260(4):959--997, 2011.
	
	\bibitem{chen2012second}
	X.~Chen.
	\newblock Second order corrections to mean field evolution for weakly
	interacting bosons in the case of three-body interactions.
	\newblock {\em Archive for Rational Mechanics and Analysis}, 203(2):455--497,
	2012.
	
	\bibitem{choozawa}
	Y.~Cho and T.~Ozawa.
	\newblock Small data scattering of {H}artree type fractional {S}chrödinger
	equations in dimension 2 and 3.
	\newblock {\em Journal of the Korean Mathematical Society}, 55(2), 2018.
	
	\bibitem{cuccagnageorgievvisciglia}
	S.~Cuccagna, V.~Georgiev, and N.~Visciglia.
	\newblock Decay and scattering of small solutions of pure power {NLS} in
	$\mathbb{R}$ with $p>3$ and with a potential.
	\newblock {arXiv:1209.5863}, 2012.
	
	\bibitem{ketterle}
	K.~B. Davis, M.-O. Mewes, M.~R. Andrews, N.~J. van Druten, D.~S. Durfee, D.~M.
	Kurn, and W.~Ketterle.
	\newblock {B}ose-{E}instein condensation in a gas of sodium atoms.
	\newblock {\em Physical Review Letters}, 75(22):3969--3973, 1995.
	
	\bibitem{de2019mean}
	G.~de~Oliveira and A.~Michelangeli.
	\newblock Mean-field dynamics for mixture condensates via fock space methods.
	\newblock {\em Reviews in Mathematical Physics}, 31(08):1950027, 2019.
	
	\bibitem{deiftzhou}
	P.~Deift and X.~Zhou.
	\newblock Long‐time asymptotics for solutions of the {NLS} equation with
	initial data in a weighted sobolev space.
	\newblock {\em Communications on Pure and Applied Mathematics},
	56(8):1029--1077, 2003.
	
	\bibitem{Dietze2021}
	C.~Dietze.
	\newblock Dispersive estimates for nonlinear {S}chr{\"o}dinger equations with
	external potentials.
	\newblock {\em J. Math. Phys.}, 62(11):111502, 2021.
	
	\bibitem{dimonte2020some}
	D.~Dimonte, M.~Falconi, and A.~Olgiati.
	\newblock On some rigorous aspects of fragmented condensation.
	\newblock {\em Nonlinearity}, 34(1):1, 2020.
	
	\bibitem{digi2021}
	D.~Dimonte and E.~L. Giacomelli.
	\newblock On {B}ose-{E}instein condensates in the {T}homas-{F}ermi regime.
	\newblock {arXiv:2112.02343}, 2021.
	
	\bibitem{duyckaertsholmerroudenko}
	T.~Duyckaerts, J.~Holmer, and S.~Roudenko.
	\newblock Scattering for the non-radial 3d cubic nonlinear {S}chrödinger
	equation.
	\newblock {arXiv:0710.3630}, 2007.
	
	\bibitem{ei1925}
	{E}instein.
	\newblock Quantentheorie des einatomigen idealen gases.
	\newblock {\em Sitzungsberichte der Preussischen Akademie der Wissenschaften},
	1(3), 1925.
	
	\bibitem{ElgartSchlein2007}
	A.~Elgart and B.~Schlein.
	\newblock Mean field dynamics of boson stars.
	\newblock {\em Comm. Pure Appl. Math.}, 60(4):500--545, 2007.
	
	\bibitem{ErdosSchlein2009}
	L.~Erd{\H{o}}s and B.~Schlein.
	\newblock Quantum dynamics with mean field interactions: a new approach.
	\newblock {\em J. Stat. Phys.}, 134(5-6):859--870, 2009.
	
	\bibitem{ErdosSchleinYau2007}
	L.~Erd{\H{o}}s, B.~Schlein, and H.-T. Yau.
	\newblock Derivation of the cubic non-linear {S}chr{\"o}dinger equation from
	quantum dynamics of many-body systems.
	\newblock {\em Inventiones mathematicae}, 167(3):515--614, 2007.
	
	\bibitem{ErdosSchleinYau2009}
	L.~Erd{\H{o}}s, B.~Schlein, and H.-T. Yau.
	\newblock Rigorous derivation of the {G}ross-{P}itaevskii equation with a large
	interaction potential.
	\newblock {\em Journal of the American Mathematical Society}, 22(4):1099--1156,
	2009.
	
	\bibitem{ErdosYau2001}
	L.~Erd{\H{o}}s and H.-T. Yau.
	\newblock Derivation of the nonlinear {S}chr\"odinger equation from a many body
	{C}oulomb system.
	\newblock {\em Adv. Theor. Math. Phys.}, 5(6):1169--1205, 2001.
	
	\bibitem{germainpusaterirousset}
	P.~Germain, F.~Pusateri, and F.~Rousset.
	\newblock The nonlinear {S}chrödinger equation with a potential.
	\newblock {\em Annales de l'Institut Henri Poincaré. Analyse non linéaire},
	35(6):1477--1530, 2018.
	
	\bibitem{ginibreozawa}
	J.~Ginibre and T.~Ozawa.
	\newblock Long range scattering for nonlinear {S}chrödinger and {H}artree
	equations in space dimension $n\geq 2$.
	\newblock {\em Communications in Mathematical Physics}, 151(3), 1993.
	
	\bibitem{GinibreVelo1979_1}
	J.~Ginibre and G.~Velo.
	\newblock The classical field limit of scattering theory for nonrelativistic
	many-boson systems. {I}.
	\newblock {\em Comm. Math. Phys.}, 66(1):37--76, 1979.
	
	\bibitem{GinibreVelo1979_2}
	J.~Ginibre and G.~Velo.
	\newblock The classical field limit of scattering theory for nonrelativistic
	many-boson systems. {II}.
	\newblock {\em Comm. Math. Phys.}, 68(1):45--68, 1979.
	
	\bibitem{ginibrevelononlocal}
	J.~Ginibre and G.~Velo.
	\newblock On a class of non linear {S}chrödinger equations with non local
	interaction.
	\newblock {\em Mathematische Zeit{S}chrift}, 170(2):109--136, 1980.
	
	\bibitem{ginibrevelotimedecaykleingordonnls}
	J.~Ginibre and G.~Velo.
	\newblock Time decay of finite energy solutions of the non linear
	{K}lein-{G}ordon and {S}chrödinger equations.
	\newblock {\em Annales de l'Institut Henri Poincaré. Physique théorique},
	43(4):399--442, 1985.
	
	\bibitem{ginibreveloscatteringhartree}
	J.~Ginibre and G.~Velo.
	\newblock Scattering theory in the energy space for a class of {H}artree
	equations.
	\newblock {arXiv:math/9809183}, 1998.
	
	\bibitem{GinibreVelo1998}
	J.~Ginibre and G.~Velo.
	\newblock {\em Scattering theory in the energy space for a class of {H}artree
		equations}, volume 263 of {\em Contemp. Math.}
	\newblock Amer. Math. Soc., Providence, RI, 2000.
	
	\bibitem{GrillakisMachedon13}
	M.~Grillakis and M.~Machedon.
	\newblock Pair excitations and the mean field approximation of interacting
	bosons, {I}.
	\newblock {\em Comm. Math. Phys.}, 324(2):601--636, 2013.
	
	\bibitem{GrillakisMachedon17}
	M.~Grillakis and M.~Machedon.
	\newblock Pair excitations and the mean field approximation of interacting
	bosons, {II}.
	\newblock {\em Commun. Partial Differ. Equ.}, 42(1):24--67, 2017.
	
	\bibitem{GrillakisMachedonMargetis2010}
	M.~G. Grillakis, M.~Machedon, and D.~Margetis.
	\newblock Second-order corrections to mean field evolution of weakly
	interacting bosons. {I}.
	\newblock {\em Comm. Math. Phys.}, 294(1):273--301, 2010.
	
	\bibitem{GrillakisMachedonMargetis2011}
	M.~G. Grillakis, M.~Machedon, and D.~Margetis.
	\newblock Second-order corrections to mean field evolution of weakly
	interacting bosons. {II}.
	\newblock {\em Adv. Math.}, 228(3):1788--1815, 2011.
	
	\bibitem{gross}
	E.~P. Gross.
	\newblock Structure of a quantized vortex in boson systems.
	\newblock {\em Nuovo Cimento}, 20:454--477, 1961.
	
	\bibitem{HayashiNaumkin98}
	N.~Hayashi and P.~I. Naumkin.
	\newblock Asymptotics for large time of solutions to the nonlinear
	{S}chr\"odinger and {H}artree equations.
	\newblock {\em Amer. J. Math.}, 120(2):369--389, 1998.
	
	\bibitem{hayashitsutsumi}
	N.~Hayashi and Y.~Tsutsumi.
	\newblock Scattering theory for {H}artree type equations.
	\newblock {\em Annales de l’Institut Henri Poincaré. Physique théorique},
	1987.
	
	\bibitem{Hepp1974}
	K.~Hepp.
	\newblock The classical limit for quantum mechanical correlation functions.
	\newblock {\em Comm. Math. Phys.}, 35:265--277, 1974.
	
	\bibitem{hong}
	Y.~Hong.
	\newblock Scattering for a nonlinear {S}chrödinger equation with a potential.
	\newblock {arXiv:1403.3944}, 2014.
	
	\bibitem{katopusateri}
	J.~Kato and F.~Pusateri.
	\newblock A new proof of long range scattering for critical nonlinear
	{S}chrödinger equations.
	\newblock {arXiv:1004.0721}, 2010.
	
	\bibitem{KnowlesPickl2010}
	A.~Knowles and P.~Pickl.
	\newblock Mean-field dynamics: singular potentials and rate of convergence.
	\newblock {\em Comm. Math. Phys.}, 298(1):101--138, 2010.
	
	\bibitem{kuz2015rate}
	E.~Kuz.
	\newblock Rate of convergence to mean field for interacting bosons.
	\newblock {\em Communications in Partial Differential Equations},
	40(10):1831--1854, 2015.
	
	\bibitem{Lee2019time}
	J.~Lee.
	\newblock On the time dependence of the rate of convergence towards {H}artree
	dynamics for interacting bosons.
	\newblock {\em J. Stat. Phys.}, 176(2):358--381, 2019.
	
	\bibitem{lee2020rate}
	J.~Lee.
	\newblock Rate of convergence towards mean-field evolution for weakly
	interacting bosons with singular three-body interactions.
	\newblock {arXiv:2006.13040}, 2020.
	
	\bibitem{Lee2021mixture}
	J.~Lee.
	\newblock Rate of convergence toward {H}artree type equations for mixture	condensates with factorized initial data.
	\newblock {\em J. Math. Phys.}, 62(9):091901, 2021.
	
	\bibitem{leemichelangeli2022}
	J.~Lee and A.~Michelangeli.
	\newblock On the characterisation of fragmented Bose-Einstein condensation and its emergent effective evolution.
	\newblock {arXiv:2211.07133}, 2022.
	
	\bibitem{Lee2013}
	J.~O. Lee.
	\newblock Rate of convergence towards semi-relativistic {H}artree dynamics.
	\newblock {\em Ann. Henri Poincar\'e}, 14(2):313--346, 2013.
	
	\bibitem{leggett}
	A.~J. Leggett.
	\newblock {B}ose-{E}instein condensation in the alkali gases: Some fundamental
	concepts.
	\newblock {\em Reviews of Modern Physics}, 73:307--356, 2001.
	
	\bibitem{lenaro2014}
	M.~Lewin, P.~T. Nam, and N.~Rougerie.
	\newblock Derivation of {}'s theory for generic mean-field {{B}ose} systems.
	\newblock {\em Adv. Math.}, 254:570--621, 2014.
	
	\bibitem{LewinNamSchlein15}
	M.~Lewin, P.~T. Nam, and B.~Schlein.
	\newblock Fluctuations around {H}artree states in the mean-field regime.
	\newblock {\em American Journal of Mathematics}, 137(6):1613--1650, 2015.
	
	\bibitem{lieb2005mathematics}
	E.~H. Lieb, R.~Seiringer, J.~P. Solovej, and J.~Yngvason.
	\newblock {\em The mathematics of the {B}ose gas and its condensation},
	volume~34.
	\newblock Springer Science \& Business Media, 2005.
	
	\bibitem{michelangeli2019ground}
	A.~Michelangeli, P.~T. Nam, and A.~Olgiati.
	\newblock Ground state energy of mixture of {B}ose gases.
	\newblock {\em Reviews in Mathematical Physics}, 31(02):1950005, 2019.
	
	\bibitem{michelangeli2017mean}
	A.~Michelangeli and A.~Olgiati.
	\newblock Mean-field quantum dynamics for a mixture of {B}ose--{E}instein
	condensates.
	\newblock {\em Analysis and Mathematical Physics}, 7(4):377--416, 2017.
	
	\bibitem{MichelangeliSchlein2012}
	A.~Michelangeli and B.~Schlein.
	\newblock Dynamical collapse of boson stars.
	\newblock {\em Comm. Math. Phys.}, 311(3):645--687, 2012.
	
	\bibitem{nakanishi}
	K.~Nakanishi.
	\newblock Global dynamics below excited solitons for the nonlinear
	{S}chrödinger equation with a potential.
	\newblock {\em Journal of the Mathematical Society of Japan},
	69(4):1353–1401, 2017.
	
	\bibitem{nana2015}
	P.~T. Nam and M.~Napi{\'o}rkowski.
	\newblock Bogoliubov correction to the mean-field dynamics of interacting
	bosons.
	\newblock {arXiv:1509.04631}, 2015.
	
	\bibitem{namnapiorkowski}
	P.~T. Nam and M.~Napiórkowski.
	\newblock A note on the validity of bogoliubov correction to mean-field
	dynamics.
	\newblock {\em Journal de mathématiques pures et appliquées},
	108(5):662--688, 2017.
	
	\bibitem{nanaritr2020}
	P.~T. Nam, M.~Napiórkowski, J.~Ricaud, and A.~Triay.
	\newblock Optimal rate of condensation for trapped bosons in the
	{G}ross-{P}itaevskii regime.
	\newblock {arXiv:2001.04364}.
	
	\bibitem{nam2021condensation}
	P.~T. Nam, J.~Ricaud, and A.~Triay.
	\newblock The condensation of a trapped dilute {B}ose gas with three-body
	interactions.
	\newblock {arXiv:2110.08195}, 2021.
	
	\bibitem{natr2021}
	P.~T. Nam and A.~Triay.
	\newblock Bogoliubov excitation spectrum of trapped {B}ose gases in the
	{G}ross-{P}itaevskii regime.
	\newblock {arXiv:2106.11949}, 2021.
	
	\bibitem{napiorkowski2021dynamics}
	M.~Napi{\'o}rkowski.
	\newblock Dynamics of interacting bosons: a compact review.
	\newblock {arXiv:2101.04594}, 2021.
	
	\bibitem{naumkinpotential}
	I.~P. Naumkin.
	\newblock Sharp asymptotic behavior of solutions for cubic nonlinear
	{S}chrödinger equations with a potential.
	\newblock {\em Journal of Mathematical Physics}, 57(5), 2016.
	
	\bibitem{naumkinexceptional}
	I.~P. Naumkin.
	\newblock Nonlinear {S}chrödinger equations with exceptional potentials.
	\newblock {\em Journal of Differential Equations}, 265(9):4575--4631, 2018.
	
	\bibitem{ozawa}
	T.~Ozawa.
	\newblock Long range scattering for nonlinear {S}chrödinger equations in one
	space dimension.
	\newblock {\em Communications in Mathematical Physics}, 139(3):479--493, 1991.
	
	\bibitem{pickl2008gphtr}
	P.~Pickl.
	\newblock On the time dependent gross pitaevskii- and {H}artree equation.
	\newblock {arXiv:0808.1178 }, 2008.
	
	\bibitem{Pickl2011}
	P.~Pickl.
	\newblock A simple derivation of mean field limits for quantum systems.
	\newblock {\em Lett. Math. Phys.}, 97(2):151--164, 2011.
	
	\bibitem{Pickl2015}
	P.~Pickl.
	\newblock Derivation of the time dependent {G}ross-{P}itaevskii equation with
	external fields.
	\newblock {\em Rev. Math. Phys.}, 27(1):1550003, 45, 2015.
	
	\bibitem{pitaevskii}
	L.~Pitaevskii.
	\newblock Vortex lines in an imperfect {B}ose gas.
	\newblock {\em Journal of Experimental and Theoretical Physics},
	13(2):451--454, 1961.
	
	\bibitem{pomeaurica}
	Y.~Pomeau and S.~Rica.
	\newblock Model of superflow with rotons.
	\newblock {\em Physical Review Letters}, 71(2):247--250, 1993.
	
	\bibitem{pusaterisoffer}
	F.~Pusateri and A.~Soffer.
	\newblock Bilinear estimates in the presence of a large potential and a
	critical {NLS} in 3d.
	\newblock {arXiv:2003.00312}, 2020.
	
	\bibitem{rs}
	I.~Rodnianski and W.~Schlag.
	\newblock Time decay for solutions of {S}chrödinger equations with rough and
	time-dependent potentials.
	\newblock {\em Interventiones Mathematicae}, 155(3):451--513, 2004.
	
	\bibitem{RodnianskiSchlein2009}
	I.~Rodnianski and B.~Schlein.
	\newblock Quantum fluctuations and rate of convergence towards mean field
	dynamics.
	\newblock {\em Comm. Math. Phys.}, 291(1):31--61, 2009.
	
	\bibitem{Spohn1980}
	H.~Spohn.
	\newblock Kinetic equations from {H}amiltonian dynamics: {M}arkovian limits.
	\newblock {\em Rev. Modern Phys.}, 52(3):569--615, 1980.
	
\end{thebibliography}
\end{document}